\documentclass[pra, twocolumn, floatfix, superscriptaddress, longbibliography, groupedaddress]{revtex4-1}

\usepackage{amsmath, amssymb}
\usepackage[dvipsnames]{xcolor}
\usepackage{color, graphicx}
\usepackage{hyperref}

\newcommand{\ket}[1]{{| #1 \rangle}}

\newcommand{\expect}[1]{{\langle #1 \rangle}}
\newcommand{\curlyL}{{\mathcal{L}}}
\newcommand{\bb}[1]{{\boldsymbol #1}}

\begin{document}

\title{Electron vortex beams in non-uniform magnetic fields}

\author{Abhijeet Melkani}
\email{amelkani@uoregon.edu}

\author{S. J. van Enk}
\email{svanenk@uoregon.edu}

\affiliation{Department of Physics and  Oregon  Center  for Optical, Molecular \&  Quantum  Sciences, University of Oregon, Eugene, Oregon, USA}

\date{\today}

\begin{abstract}

We consider the quantum theory of paraxial non-relativistic electron beams in non-uniform magnetic fields, such as the Glaser field. We find the wave function of an electron from such a beam and show that it is a joint eigenstate of two ($z$-dependent) commuting gauge-independent operators. This generalized Laguerre-Gaussian vortex beam has a phase that is shown to consist of two parts, each being proportional to the eigenvalue of one of the two conserved operators and each having different symmetries. We also describe the dynamics of the angular momentum and cross-sectional area of any mode and how a varying magnetic field can split a mode into a superposition of modes. By a suitable change in frame of reference all of our analysis also applies to an electron in a quantum Hall system with a time-dependent magnetic field.

\end{abstract}

\maketitle

\section{Introduction}
The field of electron optics \cite{hawkes1996book, grivet2013book} was pioneered by Glaser      \cite{glaser1941,glaser2013book}, based on similarities with light optics and the possibility of using electromagnetic fields as electron lenses. New life was injected into the field in the 21st century \cite{bliokh2007semiclassical, verbeeck2010production, uchida2010generation, mcmorran2011angularMomentum} with the production and exploitation of electron vortex beams. Such beams, like their optical counterparts, are hollow and can carry a large amount of quantized angular momentum in the direction of their propagation.

In transmission electron microscopy (TEM) vortex beams can be used to increase the resolution of the microscope to the atomic scale \cite{tamburini2006rayleigh, verbeeck2011atomicScale, rusz2016magnetic, schattschneider2012spinPolarized} and to probe chirality and magnetic dichroism in specimens \cite{lloyd2012momentumTransfer, schattschneider2013comment, rusz2014dichroism, yuan2013chiralSpecific, pohl2015measureEMCD, schattschneider2014dichroismFeasible, schachinger2017EMCD,edstrom2016scatteringPRL, edstrom2016scattering}. The interaction of their orbital angular momentum degrees of freedom with external magnetic fields gives rise to interesting dynamics \cite{gallatin2012propagation, greenshields2012vacuum, karimi2012conversion, littlejohn1993neutral} which endows them with information-rich phase structures \cite{aharanov1992berry, guzzinati2013observation, lubk2013transportPhase, allen2001phaseRetrieval}.  Electron vortex beams can also be used to study fundamental quantum-mechanical phenomena \cite{beche2014magnetic, ivanov2016double} including imaging Landau states (previously hidden in condensed-matter systems) \cite{schattschneider2014imaging}, performing Stern-Gerlach-like experiments \cite{batelaan1997sternGerlach, gallup2001sternGerlach, harvey2017sternExperimental}, and achieving spin-filtering applications \cite{schattschneider2017spinPolarization, karimi2014spinPolarized, grillo2013spinDevice}.

We focus here on the theoretical description of paraxial electron vortex beams, propagating along the $z$ direction in a non-uniform magnetic field pointing predominantly in the direction of propagation. The magnetic field is non-uniform in that its z-component is a function of z. In each transverse plane $z={\rm constant}$, the electron's spatial wave function can be viewed as living in a Hilbert space of square-integrable functions on the two-dimensional $x,y$ plane, and physical quantities are then represented as self-adjoint operators acting on that Hilbert space. Such an operator description based on the paraxial approximation is well known in optics \cite{stoler1981operator, vanEnk1992eigenfunction}, and we show it is a convenient description for electron beams as well yielding exact solutions of the paraxial equation.

In an inertial frame of reference travelling with the classical electron's velocity $v$ along the z axis the electron is confined to a {\em stationary} plane and subject to a {\em time-dependent} magnetic field. Our solution thus also directly applies to a time-dependent quantum Hall system. We also ensure that the physical quantities we use to describe the Hall effect are gauge-invariant and behave well in the limit of zero magnetic field.

In particular, this allows us to answer the following questions. Light waves whose electric field dependence on the azimuthal angle $\phi$ is given by $\exp(im\phi)$
are known to posses orbital angular momentum $L_z$ in the $z$ direction equal to $m\hbar$ per photon. For a free electron the same is true, but for an electron in a nonzero magnetic field both $L_z$ and the local phase of its wave function are gauge dependent. So, what does one measure when one forms an image of the electron's wave function? How is the measured phase related to the physical angular momentum of the electron and the form of the applied magnetic field?

This article is structured as follows: In Section II we derive the paraxial equation for a single electron in an arbitrary $z$-dependent magnetic field  and give its exact solution.
In Section III we show there are four physical quantities whose expectation values (as functions of $z$) satisfy a {\em closed} set of differential equations. We also construct {\em two} linear combinations of those four quantities that are conserved (as functions of $z$). Those two quantities are represented by two operators whose eigenvalue equations determine a complete basis for the electron's wave function for each $z$.
In Section IV we provide explicit analytical solutions for three important cases, including the Glaser field which is a well-established model for a magnetic lens~\cite{szilagyi1998glaser}. Those example solutions show what roles the two eigenvalues (i.e., the quantum numbers) of our two operators play in the solutions (one is that of determining phase factors and their symmetries). In Section V we show how an electron may make a transition from a single mode to a superposition of modes by propagating through a region with a varying magnetic field.
Finally, in Section VI we point out connections of our theory to recent work on emulating gauge theories for charged particles by neutral atoms in laser fields, light beams in wave guides, etc.
The theoretical descriptions of these systems are, by design, identical, but what one can measure in practice varies from one system to another.

\section{The paraxial equation and its solution}

Consider an electron of charge $-e$ and mass $m_e$ in a magnetic field $\boldsymbol{B} = B(z)\boldsymbol{\hat{z}} - \frac{\rho}{2}\frac{d B(z)}{d z} \boldsymbol{\hat{\rho}}$ \footnote{The extra radial term in the magnetic field is a consequence of the fact that $\boldsymbol\nabla. \boldsymbol B$ needs to be zero.}. (The electron's spin degree of freedom is considered in Appendix \ref{Sec:Spin}.) We use the symmetric gauge, $\boldsymbol{A} = \frac{B(z)}{2}\rho \boldsymbol{\hat{\phi}}$, so that the Hamiltonian is
\begin{equation}
    H = \frac{(\boldsymbol{p} + e\boldsymbol{A})^2}{2m_e} = T_\perp - \frac{\hbar^2}{2m_e}\frac{\partial^2}{\partial z^2}.
\end{equation}
Here $T_\perp$ is the transverse kinetic energy given in terms of the mechanical momentum, $\bb{\pi} = \bb{p}+ e\bb{A}$, by $T_\perp = \frac{\pi_x^2 + \pi_y^2}{2m_e}$. The electron beam is an approximate eigenfunction of the $p_z$ operator so we look for solutions of the form
\begin{equation}
    \Psi(\rho, \phi, z) = e^{ik z}\chi(\rho, \phi, z),
\end{equation}
where $\chi(\rho, \phi, z)$ varies slowly with $z$  (i.e., $|\frac{\partial^2\chi}{\partial z^2}| \ll k | \frac{\partial \chi}{\partial z}|$). Hence we obtain
\begin{equation}\label{paraxial}
    \frac{\partial^2\Psi}{\partial z^2} \approx e^{ikz}\bigg( 2ik \frac{\partial \chi}{\partial z} - k^2 \chi\bigg).
\end{equation}
The time-independent Schr\"odinger equation, $H\Psi = E\Psi$, can then, with the help of the paraxial approximation above, be written as
\begin{equation}\label{schrodinger}
     i\hbar v \frac{\partial \chi}{\partial z} = ( T_\perp - \Delta E ) \chi = H_{\text{eff}} \chi.
\end{equation}
Here $v = \frac{\hbar k}{m_e}$ is the velocity of the electron along the z-axis and $\Delta E = E - \frac{k^2\hbar^2}{2m_e}$. We can get rid of the constant $\Delta E$ by the redefinition $\chi \to e^{i\frac{\Delta E}{\hbar v}z} \chi$. This is equivalent, within the paraxial approximation, to setting $\Delta E$ to zero by adjusting $k$.

(Note that if one replace $z/v$ with time $t$ then the above equation is the time-dependent Schr\"odinger equation for an electron constrained to a plane and subject to the time-varying electromagnetic field $\bb A(t)$. This mapping to the time-dependent quantum Hall system and other systems will be discussed in Sec. \ref{Sec:connections}.)

Now we denote the Larmor frequency, which governs the precession of the electron in a magnetic field, by $\Omega(z) = \frac{eB(z)}{2m_e}$. The transverse kinetic energy is then
\begin{equation}\label{transverseKinetic}
T_\perp = \frac{p_x^2 + p_y^2}{2m_e} + \frac{m_e \Omega(z)^2\rho^2}{2} +\Omega(z)L_z,
\end{equation}
where $L_z$ is the z-component of the canonical angular momentum operator, $L_z = (\bb r \times \bb p)_z = -i\hbar \frac{\partial}{\partial \phi}$. This is not to be confused with the gauge-invariant mechanical angular momentum of the electron, $(\bb r \times \bb \pi)_z := \curlyL_z$ \cite{kitadono2020guidingCenter, vanEnk2020angular, wakamatsu2020gauge}. In the symmetric gauge the latter can be written as
\begin{equation}\label{defineMechAngular}
\curlyL_z  = L_z + m_e\Omega(z)\rho^2.
\end{equation}

\begin{table*}[htb]
	\begin{tabular}{c| c| c| c } 
	\hline
	\phantom{a} $\hat{O}$ \phantom{a} & \phantom{aaaaa} Physical quantity described \phantom{aaaaa}& Definition \phantom{a} & Form under circular gauge, $\boldsymbol{A} = \frac{B(z)}{2}\rho \boldsymbol{\hat{\phi}}$\\ [0.5ex]
	\hline
	$\boldsymbol \pi$ & Particle's linear momentum & $\boldsymbol p - q \boldsymbol A$ & \phantom{a} $-i\hbar\frac{\partial}{\partial \rho}\boldsymbol{\hat \rho} - \bigg (\frac{qB(z)}{2}\rho +i\hbar \frac{1}{\rho}\frac{\partial}{\partial \phi}\bigg )\boldsymbol{\hat \phi} -i\hbar \frac{\partial}{\partial z}\boldsymbol{\hat z}$\\ 
	$\boldsymbol \curlyL$ & Particle's angular momentum & $\boldsymbol r \times \boldsymbol \pi$ & $\boldsymbol r \times \boldsymbol \pi$\\
	$\boldsymbol{\curlyL}^{EM}$ & \phantom{a} EM field's angular momentum \phantom{a} & $q\boldsymbol r \times \boldsymbol A^\perp$ & $q\boldsymbol r \times \boldsymbol A = \boldsymbol r \times \bb p - \boldsymbol r \times \boldsymbol \pi$\\
	$I_z$ & Moment of inertia about the z-axis & $ m_e\boldsymbol{r}_\perp. \boldsymbol{r}_\perp$ & $m_e\rho^2$\\
	$G_\perp$ & $\frac{1}{2m_e} \frac{d}{dt} I_z$ & \phantom{a} $\frac{\boldsymbol{r}_\perp.\boldsymbol{\pi}_\perp + \boldsymbol{\pi}_\perp.\boldsymbol{r}_\perp}{2} = \boldsymbol{r}_\perp.\boldsymbol{\pi}_\perp - i\hbar$ \phantom{a}& $-i\hbar\rho \frac{\partial}{\partial\rho}- i\hbar $\\
	\hline
	\end{tabular}
	\caption{\label{Table:gaugeInvariantOperators} A table of gauge-invariant operators for a particle with mass $m_e$ and charge $q$. In the definition of $\boldsymbol{\curlyL}^{EM}$, $\boldsymbol A^\perp$  refers to the transverse component of $\boldsymbol A$ which in Fourier space is orthogonal to $\bb k$ and which in position space satisfies $\boldsymbol \nabla. \boldsymbol A^\perp = 0$ \cite{cohen1997electrodynamics}. In the symmetric gauge we have $\boldsymbol \nabla. \boldsymbol A = 0$, so $\boldsymbol A = \boldsymbol A^\perp$.}
\end{table*}

One can also consider the z component of the angular momentum of the electromagnetic field, $\curlyL_z^{EM}$, which for the present system turns out to be (see Table \ref{Table:gaugeInvariantOperators}) \citep{greenshields2014conserved} 
\begin{equation}\label{defineEMAngular}
\curlyL_z^{EM} = - m_e\Omega(z)\rho^2.
\end{equation}
Equations \eqref{defineMechAngular} and \eqref{defineEMAngular} show that the z component of the total angular momentum of the system, $\curlyL_z^{tot} = \curlyL_z + \curlyL_z^{EM}$, has the same representation as $L_z$ in the chosen gauge. However while $L_z$ is conserved only in the symmetric gauge, $\curlyL_z^{tot}$, being a physical quantity, is conserved regardless of the chosen gauge. To find a second gauge-invariant quantity that is conserved as a function of $z$  we consider the operator equation
\begin{equation}\label{ehrenfest}
\frac{dO}{dz} = \frac{\partial O}{\partial z} + \frac{i}{\hbar v} [T_\perp,  O] = 0.
\end{equation}
In order to find a solution that is independent of the solution $L_z$ we first define the operator
\begin{equation}
G_\perp = \frac{\boldsymbol{r}_\perp.\boldsymbol{\pi}_\perp + \boldsymbol{\pi}_\perp.\boldsymbol{r}_\perp}{2} = -i\hbar\rho \frac{\partial}{\partial\rho} -i\hbar,
\end{equation} 
which is proportional to the rate of change of the electron's moment of inertia about the z-axis (see Table \ref{Table:gaugeInvariantOperators}).  A gauge-invariant operator that satisfies Eq. \eqref{ehrenfest}, and is thus conserved, is the Ermakov-Lewis invariant \cite{lewis1969conserved}
\begin{align}\label{D}
I(z) &= \frac{m_e}{2\hbar} \bigg[ w^2(T_\perp - \Omega \curlyL_z) - vw\dot{w}G_\perp\nonumber\\ 
&+ \bigg( \Omega^2w^2 + \frac{v^2}{2}(\dot{w}^2 + w\ddot{w}) \bigg) m_e \rho^2\bigg],
\end{align}
where $w(z)$ is any particular solution of the Ermakov-Pinney equation \cite{leach2008ermakov},
\begin{equation}\label{lensing}
\frac{4\hbar^2}{m_e^2w^4(z)} - v^2\frac{\ddot{w}(z)}{w(z)} = \Omega^2(z),
\end{equation}
where the dot signifies the derivative with respect to $z$.  Note that this equation is invariant under
three different transformations: $v\mapsto -v$, $\Omega\mapsto -\Omega$, and $z\mapsto -z$. We show later, see Eq. \eqref{rho}, that $w(z)$ is proportional to the width of the beam. Hence from now on we will refer to Eq. \eqref{lensing} as the ``lensing equation". While it is a nonlinear equation, its solutions can be written in terms of the solutions of the {\em linear} differential equation
\begin{equation}\label{linearLensing}
    v^2\ddot w(z) + \Omega^2(z) w(z) = 0,
\end{equation}
as shown in Ref.~\cite{pinney1950nonlinear}. For $I(z)$ to be Hermitian we will only consider real-valued solutions of $w(z)$.

The fact that $\curlyL_z^{tot}$ and $I(z)$ are conserved means that their eigenvalues are independent of $z$. Also since they commute with the operator $\frac{\partial}{\partial z} + \frac{i}{\hbar v}T_\perp$ their eigenstates are the same (up to a phase factor) as the solutions $\ket{\chi}$ of Eq. \eqref{schrodinger}. Therefore,
\begin{align}
\curlyL_z^{tot}\ket{\chi(z)} &= \hbar l\ket{\chi(z)},\\
I(z)\ket{\chi(z)} &= \hbar(2n+|l|+1)\ket{\chi(z)},
\end{align}
where $l$ is an integer and $n$ is a non-negative integer. We now have two new equations for the electron's wave function. Using an Ansatz inspired by the form of a vortex beam in free space~\cite{allen1999freeSpaceSolution} we get as the solution  \cite{menouar2010wavefunction}
\begin{align}\label{solutionWF}
\chi_{n,l} = \frac{N}{w} \bigg (\frac{\rho}{w} \bigg)^{|l|}L_n^{|l|}\bigg(\frac{2\rho^2}{w^2}\bigg) \exp\bigg ( -\frac{\rho^2}{w^2} + ik \frac{\rho^2\dot{w}}{2w}\bigg )e^{il\phi-i\theta(z)}.
\end{align}
Here, $w(z)$ is a solution of Eq. \eqref{lensing}, $L_n^{|l|}(x)$ is the associated Laguerre polynomial, and $N = \sqrt{\frac{2^{|l+1|}}{\pi} \frac{n!}{(n+|l|)!}}$ is the normalization constant. $\theta(z)$ has two terms, one proportional to the eigenvalue of $I(z)$ and the other to that of $\curlyL_z^{tot}$:
\begin{equation}\label{phase}
\theta(z) = (2n+|l|+1) \frac{2\hbar}{m_e}\int_0^z \frac{dz'}{v}\frac{1}{w^2(z')} + l\int_0^z \frac{dz'}{v}\Omega(z'). 
\end{equation}

The second term is invariant with respect to the transformation $(l, \Omega) \to (-l, -\Omega)$ but under $(l, \Omega) \to (l, -\Omega)$ or $(l, \Omega) \to (-l, \Omega)$ it changes sign. The first term is invariant with respect to all three transformations. This difference in symmetry between modes with opposite vorticities $\pm l$ or opposite magnetic fields $\pm \Omega$ may be used to measure these two terms separately.

The phase of the electron's wave function is, however, gauge-dependent. The experimentally measured image of the wave function is an interference pattern which depends only on a phase \emph{difference}. The gauge-independent interference term is given by $\chi_{n,l}(x,y)\chi_{\rm ref}^*(x,y)$ where $\chi_{\rm ref}$ refers to a reference beam which has either travelled a different path or had different initial conditions or quantum numbers. If the two beams have different $l$ quantum numbers, the image (interference pattern) will rotate as the beams propagate forward~\cite{guzzinati2013observation}.

The initial conditions on any mode Eq. \eqref{solutionWF} can be specified by the initial beam waist $w(z)|_{z=0}$ and the initial radius of curvature of the wave front $R(z)|_{z=0} = \frac{w(z)}{\dot{w}(z)}\bigg|_{z=0}$ \citep{bliokh2012twist}. There is then only one unique solution of Eq. \eqref{lensing} once these initial conditions are specified.

\section{Expectation values of physical observables}

While the wave function, Eq. \eqref{solutionWF}, describes all there is to know about the problem, a simpler, and admittedly limited, description may also be useful. Such a description is provided by the expectation values of physical, gauge-invariant observables.

This is not totally trivial as several of the usual physical observables employed to describe the quantum Hall effect (for example, the coordinates of the center of the cyclotron orbit) either diverge or else vanish when $B\rightarrow 0$. All our observables have a finite nonzero limit when $B\rightarrow 0$ and therefore we can show explicitly how vortex beams that start off in free space can evolve to states that are known to describe the quantum Hall effect, i.e., Landau states.

The general solution of Eq. \eqref{schrodinger} can be written as a superposition of all the modes given by Eq. \eqref{solutionWF}:
\begin{equation}\label{genSolution}
\chi(\rho, \phi, z) = \sum_{n,l} c_{n,l} \chi_{n,l}(\rho, \phi, z).
\end{equation} 
If only one of the coefficients  $c_{n,l}$ is non-zero, then the width of the beam as a function of $z$ is given by
\begin{equation}\label{rho}
 \expect{\rho^2} = 2\pi \int_0^{\infty} \rho d\rho|\chi_{n,l}|^2\rho^2 = \bigg(n + \frac{|l|}{2}\bigg) w^2(z) 
\end{equation}
and satisfies the ``lensing equation" \eqref{lensing}. Here, we changed the variable of integration from $\rho$ to $\rho/|w(z)|$ and then used the properties of the Landau wave function $\chi^{(0)}_{n,l}$ (see Eq. \eqref{landau}).

For the general case (Eq. \eqref{genSolution}) we use Eq. \eqref{ehrenfest} and take expectation values to get the following closed set of differential equations:
\begin{subequations}\label{fourclosed}
\begin{align}
    \frac{d}{dz} \expect{T_\perp} &= \dot\Omega(z)\expect{ \curlyL_z},\\
    \frac{d}{dz}\expect{ \rho^2} &= \frac{2}{m_ev}\expect{G_\perp},\\
    \frac{d}{dz}\expect{ \curlyL_z} &=  \dot\Omega(z)m_e\expect{\rho^2} + \frac{2\Omega(z)}{v}\expect{G_\perp},\\
	\frac{d}{dz}\expect{ G_\perp} &= \frac{2}{v}\bigg ( \expect{T_\perp}-\Omega(z)\expect{\curlyL_z}\bigg).
\end{align}
\end{subequations}
The closure of these differential equations can be traced to the fact that $G_\perp, \rho^2$ and $T_\perp$ are (gauge-invariant) generators of the Lie group $SU(1,1)$. The system belongs to $U(1)\times SU(1,1)$ with $\curlyL_z^{tot}$, or equivalently $L_z$, generating $U(1)$.

For an electron in a single mode, these four expectation values as functions of $z$ are fully determined by the two quantum numbers $n$ and $l$ and the values of $w(z)$ and its derivative. Explicitly,
\begin{subequations}
\begin{align}
    \expect{T_\perp} &= \hbar l \Omega + \frac{( 2n + |l|)}{2} m_e \bigg ( \frac{ v^2}{2}(\dot w^2 + w \ddot w) + \Omega^2 w^2 \bigg ),\\
    \expect{ \rho^2} &= \frac{(2n + |l|)}{2} w^2,\\
    \expect{ \curlyL_z} &= \hbar l + \frac{(2n + |l|)}{2} m_e \Omega w^2,\\
	\expect{ G_\perp} &= \frac{(2n + |l|)}{2} m_ev w \dot w.
\end{align}
\end{subequations}
In terms of these we have
\begin{align}
I(z) &= \frac{m_e}{\hbar(2n + |l|)} \bigg[ \expect{\rho^2}(T_\perp - \Omega \curlyL_z) - \frac{\expect{G_\perp}}{m_e} G_\perp\nonumber\\ 
&+ (\expect{T_\perp} - \hbar l \Omega) \rho^2\bigg]
\end{align}
for a single mode.

\section{Example solutions}

We have reduced the general problem to a linear differential equation, Eq. \eqref{linearLensing}, of one variable $w(z)$. While analytical solutions for many cases are known (see Refs. \cite{lewis1968invariant} and \cite{eliezer1976note} for example), we discuss here only three cases of the most importance to electron optics.
\subsection{Constant magnetic field}

When the magnetic field is constant, $\Omega(z) = \Omega$, the general solution of the lensing equation \eqref{lensing} can be written as
\begin{align}\label{generalForConstant}
w^2(z) =& w_0^2 \bigg[ \cos^2(\Omega z/v) + \bigg( 1 + \frac{4R_0^2}{w_0^4k^2} \bigg) \frac{v^2}{\Omega^2R_0^2} \sin^2(\Omega z/v)\nonumber\\
 &+ \frac{v}{\Omega R_0} \sin(2\Omega z/v) \bigg],
\end{align}
where $w_0 = w(z)|_{z=0}$ and $R_0 = R(z)|_{z=0} = \frac{w(z)}{\dot{w}(z)}\bigg|_{z=0}$ determine the initial conditions. With the special choice of $w_0^2 = \frac{2v}{k|\Omega|} = \frac{2\hbar}{m_e|\Omega|}$ and $R_0 \to \infty$ the solution $w(z)$ becomes constant. The corresponding wave-functions (which were first derived by Landau~\cite{landau1930original}) are eigenstates of the Hamiltonian $T_\perp$,
\begin{equation}\label{landau}
\chi^{(0)}_{n,l} = \frac{N}{w_0} \bigg (\frac{\rho}{w_0} \bigg)^{|l|}L_n^{|l|}\bigg(\frac{2\rho^2}{w_0^2}\bigg) \exp\bigg ( -\frac{\rho^2}{w_0^2}\bigg )e^{il\phi-i\theta(z)}.
\end{equation}
The phase is $\theta(z) = (2n+|l|+1)\frac{|\Omega|z}{v} + l\frac{\Omega z}{v}$.

\subsection{Free space}

The solution for free space is known to display surprisingly different behavior than that for a constant non-zero field \citep{allen1999freeSpaceSolution, bliokh2017theory}. This can also be seen in Eq. \eqref{generalForConstant} which becomes undefined when $\Omega = 0$. Taking the limit $\Omega \to 0$ gives
\begin{equation}
w^2(z) = w_0^2 \bigg[ 1 + \bigg( \frac{1}{R_0^2} + \frac{4}{w_0^4k^2} \bigg)z^2 + \frac{2z}{R_0} \bigg].
\end{equation}
It is now clear that it is impossible to have a choice of initial conditions for which  the width $w(z)$ remains non-zero constant. Therefore, the radius of curvature $R(z)$ of the beam is finite and the beam will always diffract.

The wave function of the electron is given by Eq. \eqref{solutionWF} where the phase can be written in terms of the Rayleigh diffraction length, $z_R = \frac{kw_0^2}{2}$, as
\begin{align}
    \theta(z) &= (2n+|l|+1)\bigg[  \tan^{-1}{\bigg( \frac{z}{z_R}\bigg (\frac{z_R}{R_0} + \frac{R_0}{z_R} \bigg) - \frac{z_R}{R_0} \bigg)} \nonumber\\ &+ \tan^{-1}{\bigg( \frac{z_R}{R_0} \bigg)}\bigg ].
\end{align}

\subsection{Glaser field}

The Glaser field,
\begin{equation}
    B(z) = \frac{B_0}{1 + (z- c)^2/a^2},
\end{equation}
is a very useful model for describing an electron lens. It is an approximation of the magnetic field near the axis of a single current carrying coil \cite{szilagyi1998glaser}.
The general solution of Eq. \eqref{lensing} for $\Omega(z) = \Omega_0/(1 + (z- c)^2/a^2)$ is (see Appendix \ref{Sec:Derivation} for derivation)
\begin{align}\label{Glaserlensing}
    w^2(z) &= w_c^2 \frac{a^2 + (z - c)^2}{a^2}\bigg [ \cos^2(\alpha(z)) + \frac{a}{\beta R_c}\sin(2\alpha(z)) \nonumber \\ &+\frac{a^2}{\beta^2} \bigg( \frac{1}{R_c^2} + \frac{4}{k^2w_c^4} \bigg) \sin^2(\alpha(z))  \bigg],
\end{align}
where
\begin{equation}
    \alpha(z) = \beta \arctan\bigg ( \frac{z-c}{a}\bigg),
\end{equation}
and $\beta = \sqrt{1 + a^2\Omega_0^2/v^2}$. Here the arbitrary initial conditions have been specified using both the width $w_c$ and the radius of curvature $R_c$ at $z = c$. The well-known focusing behavior of the Glaser field~\cite{szilagyi1998glaser} is demonstrated in how the width, $w(z)$, is minimized to $w_c$ at the center of the field, $z=c$. This becomes clearer when,  similar to the case of the constant magnetic field, we get rid of the oscillations by a special choice of $w_c = \frac{2a}{\beta k}$ and $R_c \to \infty$, which yields
\begin{equation}
w^2(z) = \frac{2}{ka}\frac{1}{\sqrt{1 + a^2\Omega_0^2/v^2}} \bigg ( (z - c)^2 + a^2 \bigg).
\end{equation}
The phase of the beam for this special case is
\begin{align}
    \theta(z) &= \bigg((2n+|l|+1)\sqrt{1 + a^2\Omega_0^2/v^2} + l\Omega_0a/v\bigg) \nonumber \\ &\times\bigg[  \tan^{-1}{\bigg (\frac{z-c}{a} \bigg )} + \tan^{-1}{\bigg( \frac{c}{a} \bigg)}\bigg ].
\end{align}

\section{Mode splitting}

Analogous to the analysis of thin lenses in light optics we may interested in the problem of the asymptotic final states obtained after the beam passes through an optical apparatus. We assume that the magnetic field is constant for the region $z < z_i$, varies in the region $z_i < z < z_f$, and then finally becomes constant again in the region $z > z_f$.

We take the initial wave function to have $w(z) = w_i = \sqrt{\frac{2\hbar}{m_e|\Omega_i|}}$ (and therefore $R(z) = \frac{w(z)}{\dot w(z)} \to \infty$). This wave function corresponds to the one given in Eq. \eqref{landau} and we will call such a wave function a pure Landau mode. After passing through the optical apparatus ($z_i < z < z_f$) if the final wave function has $w(z) = \sqrt{\frac{2\hbar}{m_e|\Omega_f|}}$ for $z > z_f$ then the Landau mode is unchanged and this is similar to adiabatic following.

However, if $w(z)$ takes the more general form (Eq. \eqref{generalForConstant}), then the asymptotic wave function at $z \gg z_f$ will no longer be an eigenstate of the z-independent $H_{\text{eff}}(z_f)$ (even though it will still satisfy Eq. \eqref{schrodinger}). This wave function can then be expressed as a superposition of  Landau modes. In Figure 1 we show such mode-splitting for a ramp magnetic field (analytical solutions for which are known \cite{lewis1968invariant} in terms of Bessel functions).

\begin{figure}[htb]
	\includegraphics[width=\columnwidth]{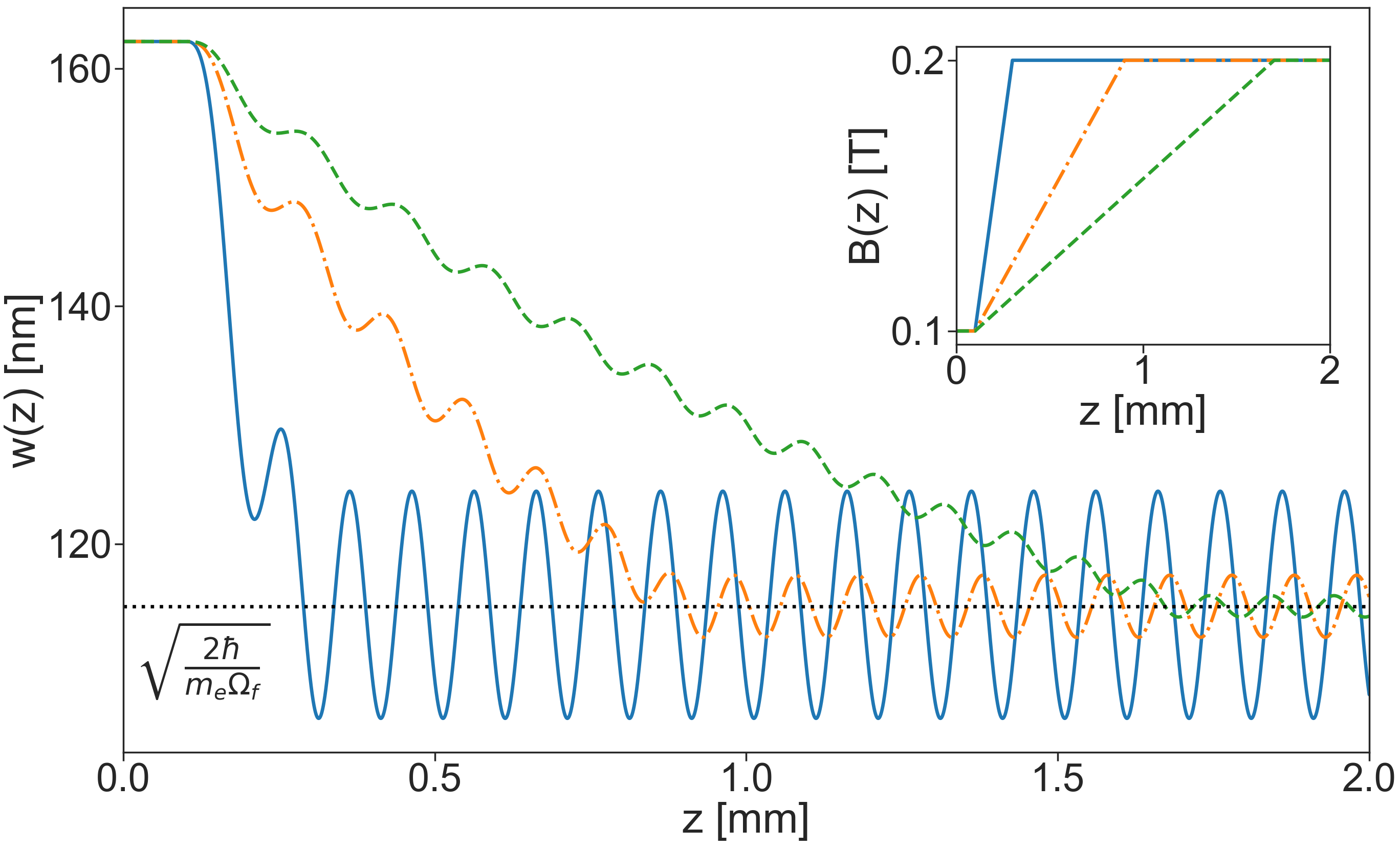}
	\caption[]{\label{Fig:splitting} 
	Mode splitting by a ramp potential: A pure Landau mode, Eq. \eqref{landau}, in a uniform magnetic field, $B = 0.1$T ($\Omega_i \approx 8.8$ GHz), travels towards the positive $z$-axis with speed $0.02$ c. Since it is a pure Landau mode $w(z) = \sqrt{\frac{2\hbar}{m_e\Omega_i}}$ is constant.
	The beam is then subject to a ramp magnetic field (see inset) in the region $z > 0.1$ mm until the magnetic field becomes $B = 0.2$ T ($\Omega_f \approx 17.6$ GHz) . The width of the beam, $w(z)$, in the constant magnetic field region has now picked up an oscillating behaviour and hence cannot be a pure Landau mode (which has a constant $w(z) = \sqrt{\frac{2\hbar}{m_e\Omega_f}}$ shown by the black dotted line).  If the slope of the ramp is gradual (green dashed curve) we see an effective adiabatic behaviour --- the oscillations are relatively small. For the extreme case of slope $\to \infty$ the exact solution can be calculated using Eq. \eqref{convenient} with $g_i(z) = g_f(z) = 1$. 
	}
\end{figure}

While the general problem will require an exact solution of the lensing equation \eqref{lensing} we here follow Ref.~\cite{lewis1969conserved} to provide an argument for why mode splitting can be expected. We first assume a convenient representation of the solution $w(z)$,
\begin{align}\label{convenient}
w^2(z) = 
\begin{cases} \frac{2\hbar}{m_e|\Omega_i|} g_i(z) &\mbox{if } z < 0 \\
w^2_{z>z_f}(z)g_f(z) & \mbox{if } z > 0 \end{cases} \end{align}
where $w^2_{z>z_f}(z)$ is given by Eq. \eqref{generalForConstant} with arbitrary constants $w_0$ and $R_0$ (these are no longer to be interpreted as the width and radius of the beam at $z=0$). Here $g_i$ and $g_f$ are continuous functions with continuous derivatives such that $g_i(z) = 1$ for $z < z_i$ and $g_f(z) = 1$ for $z > z_f$.

Since the wave function and its derivative need to be continuous we demand that $w(z)$ and $\dot w(z)$ be continuous functions. However $\Omega(z)$, and therefore $\ddot w(z)$, may be allowed to have a discontinuity at $z = 0$. The Hamiltonian is therefore only piece-wise constant.

Enforcing the continuity of $w(z)$ and $\dot w(z)$ at $z=0$ gives
\begin{equation}
    w_0^2 = \frac{2\hbar}{m_e|\Omega_i|} \frac{g_i(0)}{g_f(0)} \quad \text{and} \quad R_0 = \frac{2}{\frac{\dot{g}_i(0)}{g_i(0)} - \frac{\dot{g}_f(0)}{g_f(0)}}.
\end{equation}
The mode splits if
\begin{equation}
    w_0^2 = \frac{2\hbar}{m_e|\Omega_i|} \frac{g_i(0)}{g_f(0)} \neq \frac{2\hbar}{m_e|\Omega_f|}
\end{equation}
or equivalently $\frac{g_i(0)}{|\Omega_i|} \neq \frac{g_f(0)}{|\Omega_f|}$.

If one assumes that the magnetic field varies slowly such that $ |v\dot{\Omega}(z)| \ll \Omega^2(z)$, (while staying within the paraxial regime of Eq.\eqref{paraxial}, $|\frac{\partial^2\chi}{\partial z^2}| \ll \frac{m_ev}{\hbar} | \frac{\partial \chi}{\partial z}|$), then the mode does not split~\cite{lewis1969conserved}.

\section{Summary and Outlook}\label{Sec:connections}

In summary, we have derived the wave function of a paraxial electron beam moving in a non-uniform ($z$-dependent) magnetic field. We found two operators, one angular momentum and the other denoted by $I(z)$ of Eq.~(\ref{D}), that are conserved as functions of $z$. Their eigenvalues are constant and their joint eigenstates form a complete basis for any transverse plane, which generalize the well-known Laguerre-Gaussian beams. We also constructed a set of four operators whose expectation values as functions of $z$ form a closed set of (differential) equations, Eqs.~(\ref{fourclosed}). These four quantities can then be expressed in terms of the two eigenvalues and the width $w(z)$ of the beam and its derivative $dw/dz$. Such a simple description is expected to have applications in lensing and directing beams. The phase of the beam is easily expressed as a function of $w(z)$, the magnetic field $B(z)$ and the two eigenvalues. This may be used for better image analysis of interference patterns in microscopy.  Equation (\ref{lensing}) for the width $w(z)$ can be solved analytically, and we provided complete solutions for the free field, a constant magnetic field, and the Glaser field, highlighting the relations between them. Finally we also show how a Landau mode after passing through an optical apparatus can split into a superposition of Landau modes hinting at possible applications in production of modes with high angular momenta.

In the past few years many experimental applications have been achieved using electron vortex beams in magnetic fields. Their description could benefit from our theory especially in cases where previous analysis was done either under assumptions of adiabaticity  \cite{gallup2001sternGerlach, littlejohn1993neutral} or semi-classical dynamics \cite{batelaan1997sternGerlach, bliokh2007semiclassical}.

Finally, as previously noted, the system of an electron constrained to a plane and subject to a time-dependent potential $\boldsymbol{A} = \frac{B(t)}{2}\rho \boldsymbol{\hat{\phi}}$ leads to exactly the same equations (with $z/v$ replaced by $t$) as our equations \eqref{schrodinger} and \eqref{transverseKinetic}, namely,
\begin{equation}
     i\hbar \frac{\partial \chi}{\partial t} = \bigg (\frac{p_x^2 + p_y^2}{2m_e} + \frac{\mu \Omega(t)^2\rho^2}{2} +\Omega(t)L_z \bigg ) \chi,
\end{equation}
with $\Omega(t) = \frac{eB(t)}{2m_e}$. This corresponds to a time-varying magnetic field, $\bb B(t) = \bb \nabla \times \bb A= B(t)\boldsymbol{\hat{z}}$ {\em and} an electric field $\bb E = -\frac{\partial \bb A}{\partial t} = - \frac{B'(t)\rho}{2} \boldsymbol{\hat{\phi}}$ due to Faraday's law. All of our analysis then also applies to this system of an electron in a quantum Hall system with time-dependent magnetic field after the mapping $z/v \to t$.

In fact, a similar equation can also be found in photonic topological insulators (see Eq. 2 of \cite{rechtsman2013photonic}). Here the ``vector potential" $\bb A$ does not represent an actual magnetic field but is an ``artificial gauge field'' that arises from being in a frame that rotates along with a helical waveguide.

Such artificial gauge fields are not just used in photonics to guide light beams \cite{lumer2019lightGuiding} but may also be used to generate sonic Landau levels in metamaterials \citep{abbaszadeh2017sonicLandau} (the gauge field being generated by non-uniform strains).
For neutral atoms as well, it has been shown that laser fields can be used to mimic the dynamics of charged particles in a gauge field
\cite{dalibard2011artificialGauge}.

We expect our theoretical advances to be of relevance in each of these systems. Having different physical realizations of the same equation has the added benefit of being able to measure different quantities.

\section{Acknowledgements}

We thank Ben McMorran, Will Parker, and Jayson Paulose for useful discussion.

\bibliography{references}

\begin{thebibliography}{62}%
\makeatletter
\providecommand \@ifxundefined [1]{%
 \@ifx{#1\undefined}
}%
\providecommand \@ifnum [1]{%
 \ifnum #1\expandafter \@firstoftwo
 \else \expandafter \@secondoftwo
 \fi
}%
\providecommand \@ifx [1]{%
 \ifx #1\expandafter \@firstoftwo
 \else \expandafter \@secondoftwo
 \fi
}%
\providecommand \natexlab [1]{#1}%
\providecommand \enquote  [1]{``#1''}%
\providecommand \bibnamefont  [1]{#1}%
\providecommand \bibfnamefont [1]{#1}%
\providecommand \citenamefont [1]{#1}%
\providecommand \href@noop [0]{\@secondoftwo}%
\providecommand \href [0]{\begingroup \@sanitize@url \@href}%
\providecommand \@href[1]{\@@startlink{#1}\@@href}%
\providecommand \@@href[1]{\endgroup#1\@@endlink}%
\providecommand \@sanitize@url [0]{\catcode `\\12\catcode `\$12\catcode
  `\&12\catcode `\#12\catcode `\^12\catcode `\_12\catcode `\%12\relax}%
\providecommand \@@startlink[1]{}%
\providecommand \@@endlink[0]{}%
\providecommand \url  [0]{\begingroup\@sanitize@url \@url }%
\providecommand \@url [1]{\endgroup\@href {#1}{\urlprefix }}%
\providecommand \urlprefix  [0]{URL }%
\providecommand \Eprint [0]{\href }%
\providecommand \doibase [0]{http://dx.doi.org/}%
\providecommand \selectlanguage [0]{\@gobble}%
\providecommand \bibinfo  [0]{\@secondoftwo}%
\providecommand \bibfield  [0]{\@secondoftwo}%
\providecommand \translation [1]{[#1]}%
\providecommand \BibitemOpen [0]{}%
\providecommand \bibitemStop [0]{}%
\providecommand \bibitemNoStop [0]{.\EOS\space}%
\providecommand \EOS [0]{\spacefactor3000\relax}%
\providecommand \BibitemShut  [1]{\csname bibitem#1\endcsname}%
\let\auto@bib@innerbib\@empty
\bibitem [{\citenamefont {Hawkes}\ and\ \citenamefont
  {Kasper}(2018)}]{hawkes1996book}%
  \BibitemOpen
  \bibfield  {author} {\bibinfo {author} {\bibfnamefont {P.}~\bibnamefont
  {Hawkes}}\ and\ \bibinfo {author} {\bibfnamefont {E.}~\bibnamefont
  {Kasper}},\ }\href {\doibase 10.1016/C2015-0-06652-7} {\emph {\bibinfo
  {title} {Principles of electron optics (Second Edition)}}}\ (\bibinfo
  {publisher} {Elsevier},\ \bibinfo {year} {2018})\BibitemShut {NoStop}%
\bibitem [{\citenamefont {Grivet}(1972)}]{grivet2013book}%
  \BibitemOpen
  \bibfield  {author} {\bibinfo {author} {\bibfnamefont {P.}~\bibnamefont
  {Grivet}},\ }\href {\doibase 10.1016/C2013-0-02400-0} {\emph {\bibinfo
  {title} {Electron optics (Second Edition)}}}\ (\bibinfo  {publisher}
  {Elsevier},\ \bibinfo {year} {1972})\BibitemShut {NoStop}%
\bibitem [{\citenamefont {Glaser}(1941)}]{glaser1941}%
  \BibitemOpen
  \bibfield  {author} {\bibinfo {author} {\bibfnamefont {W.}~\bibnamefont
  {Glaser}},\ }\bibfield  {title} {\enquote {\bibinfo {title} {Strenge
  {B}erechnung magnetischer {L}insen der {F}eldform ${H}=
  {H}_0/(1+(z/a))^2$},}\ }\href {\doibase 10.1007/BF01676330} {\bibfield
  {journal} {\bibinfo  {journal} {Zeitschrift f{\"u}r Physik}\ }\textbf
  {\bibinfo {volume} {117}},\ \bibinfo {pages} {285--315} (\bibinfo {year}
  {1941})}\BibitemShut {NoStop}%
\bibitem [{\citenamefont {Glaser}(2013)}]{glaser2013book}%
  \BibitemOpen
  \bibfield  {author} {\bibinfo {author} {\bibfnamefont {W.}~\bibnamefont
  {Glaser}},\ }\href {\doibase 10.1007/978-3-662-25699-2} {\emph {\bibinfo
  {title} {Grundlagen der Elektronenoptik}}}\ (\bibinfo  {publisher}
  {Springer-Verlag},\ \bibinfo {year} {2013})\BibitemShut {NoStop}%
\bibitem [{\citenamefont {Bliokh}\ \emph {et~al.}(2007)\citenamefont {Bliokh},
  \citenamefont {Bliokh}, \citenamefont {Savel’ev},\ and\ \citenamefont
  {Nori}}]{bliokh2007semiclassical}%
  \BibitemOpen
  \bibfield  {author} {\bibinfo {author} {\bibfnamefont {K.~Y.}\ \bibnamefont
  {Bliokh}}, \bibinfo {author} {\bibfnamefont {Y.~P.}\ \bibnamefont {Bliokh}},
  \bibinfo {author} {\bibfnamefont {S.}~\bibnamefont {Savel’ev}}, \ and\
  \bibinfo {author} {\bibfnamefont {F.}~\bibnamefont {Nori}},\ }\bibfield
  {title} {\enquote {\bibinfo {title} {Semiclassical dynamics of electron wave
  packet states with phase vortices},}\ }\href {\doibase
  10.1103/PhysRevLett.99.190404} {\bibfield  {journal} {\bibinfo  {journal}
  {Phys. Rev. Lett.}\ }\textbf {\bibinfo {volume} {99}},\ \bibinfo {pages}
  {190404} (\bibinfo {year} {2007})}\BibitemShut {NoStop}%
\bibitem [{\citenamefont {Verbeeck}\ \emph {et~al.}(2010)\citenamefont
  {Verbeeck}, \citenamefont {Tian},\ and\ \citenamefont
  {Schattschneider}}]{verbeeck2010production}%
  \BibitemOpen
  \bibfield  {author} {\bibinfo {author} {\bibfnamefont {J.}~\bibnamefont
  {Verbeeck}}, \bibinfo {author} {\bibfnamefont {H.}~\bibnamefont {Tian}}, \
  and\ \bibinfo {author} {\bibfnamefont {P.}~\bibnamefont {Schattschneider}},\
  }\bibfield  {title} {\enquote {\bibinfo {title} {Production and application
  of electron vortex beams},}\ }\href {\doibase 10.1038/nature09366} {\bibfield
   {journal} {\bibinfo  {journal} {Nature}\ }\textbf {\bibinfo {volume}
  {467}},\ \bibinfo {pages} {301--304} (\bibinfo {year} {2010})}\BibitemShut
  {NoStop}%
\bibitem [{\citenamefont {Uchida}\ and\ \citenamefont
  {Tonomura}(2010)}]{uchida2010generation}%
  \BibitemOpen
  \bibfield  {author} {\bibinfo {author} {\bibfnamefont {M.}~\bibnamefont
  {Uchida}}\ and\ \bibinfo {author} {\bibfnamefont {A.}~\bibnamefont
  {Tonomura}},\ }\bibfield  {title} {\enquote {\bibinfo {title} {Generation of
  electron beams carrying orbital angular momentum},}\ }\href {\doibase
  10.1038/nature08904} {\bibfield  {journal} {\bibinfo  {journal} {Nature}\
  }\textbf {\bibinfo {volume} {464}},\ \bibinfo {pages} {737--739} (\bibinfo
  {year} {2010})}\BibitemShut {NoStop}%
\bibitem [{\citenamefont {McMorran}\ \emph {et~al.}(2011)\citenamefont
  {McMorran}, \citenamefont {Agrawal}, \citenamefont {Anderson}, \citenamefont
  {Herzing}, \citenamefont {Lezec}, \citenamefont {McClelland},\ and\
  \citenamefont {Unguris}}]{mcmorran2011angularMomentum}%
  \BibitemOpen
  \bibfield  {author} {\bibinfo {author} {\bibfnamefont {B.~J.}\ \bibnamefont
  {McMorran}}, \bibinfo {author} {\bibfnamefont {A.}~\bibnamefont {Agrawal}},
  \bibinfo {author} {\bibfnamefont {I.~M.}\ \bibnamefont {Anderson}}, \bibinfo
  {author} {\bibfnamefont {A.~A.}\ \bibnamefont {Herzing}}, \bibinfo {author}
  {\bibfnamefont {H.~J.}\ \bibnamefont {Lezec}}, \bibinfo {author}
  {\bibfnamefont {J.~J.}\ \bibnamefont {McClelland}}, \ and\ \bibinfo {author}
  {\bibfnamefont {J.}~\bibnamefont {Unguris}},\ }\bibfield  {title} {\enquote
  {\bibinfo {title} {Electron vortex beams with high quanta of orbital angular
  momentum},}\ }\href {\doibase 10.1126/science.1198804} {\bibfield  {journal}
  {\bibinfo  {journal} {Science}\ }\textbf {\bibinfo {volume} {331}},\ \bibinfo
  {pages} {192--195} (\bibinfo {year} {2011})}\BibitemShut {NoStop}%
\bibitem [{\citenamefont {Tamburini}\ \emph {et~al.}(2006)\citenamefont
  {Tamburini}, \citenamefont {Anzolin}, \citenamefont {Umbriaco}, \citenamefont
  {Bianchini},\ and\ \citenamefont {Barbieri}}]{tamburini2006rayleigh}%
  \BibitemOpen
  \bibfield  {author} {\bibinfo {author} {\bibfnamefont {F.}~\bibnamefont
  {Tamburini}}, \bibinfo {author} {\bibfnamefont {G.}~\bibnamefont {Anzolin}},
  \bibinfo {author} {\bibfnamefont {G.}~\bibnamefont {Umbriaco}}, \bibinfo
  {author} {\bibfnamefont {A.}~\bibnamefont {Bianchini}}, \ and\ \bibinfo
  {author} {\bibfnamefont {C.}~\bibnamefont {Barbieri}},\ }\bibfield  {title}
  {\enquote {\bibinfo {title} {Overcoming the {R}ayleigh criterion limit with
  optical vortices},}\ }\href {\doibase 10.1103/PhysRevLett.97.163903}
  {\bibfield  {journal} {\bibinfo  {journal} {Phys. Rev. Lett.}\ }\textbf
  {\bibinfo {volume} {97}},\ \bibinfo {pages} {163903} (\bibinfo {year}
  {2006})}\BibitemShut {NoStop}%
\bibitem [{\citenamefont {Verbeeck}\ \emph {et~al.}(2011)\citenamefont
  {Verbeeck}, \citenamefont {Schattschneider}, \citenamefont {Lazar},
  \citenamefont {St\"oger-Pollach}, \citenamefont {L\"offler}, \citenamefont
  {Steiger-Thirsfeld},\ and\ \citenamefont
  {Van~Tendeloo}}]{verbeeck2011atomicScale}%
  \BibitemOpen
  \bibfield  {author} {\bibinfo {author} {\bibfnamefont {J.}~\bibnamefont
  {Verbeeck}}, \bibinfo {author} {\bibfnamefont {P.}~\bibnamefont
  {Schattschneider}}, \bibinfo {author} {\bibfnamefont {S.}~\bibnamefont
  {Lazar}}, \bibinfo {author} {\bibfnamefont {M.}~\bibnamefont
  {St\"oger-Pollach}}, \bibinfo {author} {\bibfnamefont {S.}~\bibnamefont
  {L\"offler}}, \bibinfo {author} {\bibfnamefont {A.}~\bibnamefont
  {Steiger-Thirsfeld}}, \ and\ \bibinfo {author} {\bibfnamefont
  {G.}~\bibnamefont {Van~Tendeloo}},\ }\bibfield  {title} {\enquote {\bibinfo
  {title} {Atomic scale electron vortices for nanoresearch},}\ }\href {\doibase
  10.1063/1.3662012} {\bibfield  {journal} {\bibinfo  {journal} {Applied
  Physics Letters}\ }\textbf {\bibinfo {volume} {99}},\ \bibinfo {pages}
  {203109} (\bibinfo {year} {2011})}\BibitemShut {NoStop}%
\bibitem [{\citenamefont {Rusz}\ \emph {et~al.}(2016)\citenamefont {Rusz},
  \citenamefont {Muto}, \citenamefont {Spiegelberg}, \citenamefont {Adam},
  \citenamefont {Tatsumi}, \citenamefont {B{\"u}rgler}, \citenamefont
  {Oppeneer},\ and\ \citenamefont {Schneider}}]{rusz2016magnetic}%
  \BibitemOpen
  \bibfield  {author} {\bibinfo {author} {\bibfnamefont {J.}~\bibnamefont
  {Rusz}}, \bibinfo {author} {\bibfnamefont {S.}~\bibnamefont {Muto}}, \bibinfo
  {author} {\bibfnamefont {J.}~\bibnamefont {Spiegelberg}}, \bibinfo {author}
  {\bibfnamefont {R.}~\bibnamefont {Adam}}, \bibinfo {author} {\bibfnamefont
  {K.}~\bibnamefont {Tatsumi}}, \bibinfo {author} {\bibfnamefont {D.~E.}\
  \bibnamefont {B{\"u}rgler}}, \bibinfo {author} {\bibfnamefont {P.~M.}\
  \bibnamefont {Oppeneer}}, \ and\ \bibinfo {author} {\bibfnamefont {C.~M.}\
  \bibnamefont {Schneider}},\ }\bibfield  {title} {\enquote {\bibinfo {title}
  {Magnetic measurements with atomic-plane resolution},}\ }\href {\doibase
  10.1038/ncomms12672} {\bibfield  {journal} {\bibinfo  {journal} {Nature
  communications}\ }\textbf {\bibinfo {volume} {7}},\ \bibinfo {pages} {1--7}
  (\bibinfo {year} {2016})}\BibitemShut {NoStop}%
\bibitem [{\citenamefont {Schattschneider}\ \emph {et~al.}(2012)\citenamefont
  {Schattschneider}, \citenamefont {Schaffer}, \citenamefont {Ennen},\ and\
  \citenamefont {Verbeeck}}]{schattschneider2012spinPolarized}%
  \BibitemOpen
  \bibfield  {author} {\bibinfo {author} {\bibfnamefont {P.}~\bibnamefont
  {Schattschneider}}, \bibinfo {author} {\bibfnamefont {B.}~\bibnamefont
  {Schaffer}}, \bibinfo {author} {\bibfnamefont {I.}~\bibnamefont {Ennen}}, \
  and\ \bibinfo {author} {\bibfnamefont {J.}~\bibnamefont {Verbeeck}},\
  }\bibfield  {title} {\enquote {\bibinfo {title} {Mapping spin-polarized
  transitions with atomic resolution},}\ }\href {\doibase
  10.1103/PhysRevB.85.134422} {\bibfield  {journal} {\bibinfo  {journal} {Phys.
  Rev. B}\ }\textbf {\bibinfo {volume} {85}},\ \bibinfo {pages} {134422}
  (\bibinfo {year} {2012})}\BibitemShut {NoStop}%
\bibitem [{\citenamefont {Lloyd}\ \emph {et~al.}(2012)\citenamefont {Lloyd},
  \citenamefont {Babiker},\ and\ \citenamefont
  {Yuan}}]{lloyd2012momentumTransfer}%
  \BibitemOpen
  \bibfield  {author} {\bibinfo {author} {\bibfnamefont {S.}~\bibnamefont
  {Lloyd}}, \bibinfo {author} {\bibfnamefont {M.}~\bibnamefont {Babiker}}, \
  and\ \bibinfo {author} {\bibfnamefont {J.}~\bibnamefont {Yuan}},\ }\bibfield
  {title} {\enquote {\bibinfo {title} {Quantized orbital angular momentum
  transfer and magnetic dichroism in the interaction of electron vortices with
  matter},}\ }\href {\doibase 10.1103/PhysRevLett.108.074802} {\bibfield
  {journal} {\bibinfo  {journal} {Phys. Rev. Lett.}\ }\textbf {\bibinfo
  {volume} {108}},\ \bibinfo {pages} {074802} (\bibinfo {year}
  {2012})}\BibitemShut {NoStop}%
\bibitem [{\citenamefont {Schattschneider}\ \emph {et~al.}(2013)\citenamefont
  {Schattschneider}, \citenamefont {L\"offler},\ and\ \citenamefont
  {Verbeeck}}]{schattschneider2013comment}%
  \BibitemOpen
  \bibfield  {author} {\bibinfo {author} {\bibfnamefont {P.}~\bibnamefont
  {Schattschneider}}, \bibinfo {author} {\bibfnamefont {S.}~\bibnamefont
  {L\"offler}}, \ and\ \bibinfo {author} {\bibfnamefont {J.}~\bibnamefont
  {Verbeeck}},\ }\bibfield  {title} {\enquote {\bibinfo {title} {Comment on
  ``{Q}uantized orbital angular momentum transfer and magnetic dichroism in the
  interaction of electron vortices with matter''},}\ }\href {\doibase
  10.1103/PhysRevLett.110.189501} {\bibfield  {journal} {\bibinfo  {journal}
  {Phys. Rev. Lett.}\ }\textbf {\bibinfo {volume} {110}},\ \bibinfo {pages}
  {189501} (\bibinfo {year} {2013})}\BibitemShut {NoStop}%
\bibitem [{\citenamefont {Rusz}\ \emph {et~al.}(2014)\citenamefont {Rusz},
  \citenamefont {Idrobo},\ and\ \citenamefont {Bhowmick}}]{rusz2014dichroism}%
  \BibitemOpen
  \bibfield  {author} {\bibinfo {author} {\bibfnamefont {J.}~\bibnamefont
  {Rusz}}, \bibinfo {author} {\bibfnamefont {J.-C.}\ \bibnamefont {Idrobo}}, \
  and\ \bibinfo {author} {\bibfnamefont {S.}~\bibnamefont {Bhowmick}},\
  }\bibfield  {title} {\enquote {\bibinfo {title} {Achieving atomic resolution
  magnetic dichroism by controlling the phase symmetry of an electron probe},}\
  }\href {\doibase 10.1103/PhysRevLett.113.145501} {\bibfield  {journal}
  {\bibinfo  {journal} {Phys. Rev. Lett.}\ }\textbf {\bibinfo {volume} {113}},\
  \bibinfo {pages} {145501} (\bibinfo {year} {2014})}\BibitemShut {NoStop}%
\bibitem [{\citenamefont {Yuan}\ \emph {et~al.}(2013)\citenamefont {Yuan},
  \citenamefont {Lloyd},\ and\ \citenamefont
  {Babiker}}]{yuan2013chiralSpecific}%
  \BibitemOpen
  \bibfield  {author} {\bibinfo {author} {\bibfnamefont {J.}~\bibnamefont
  {Yuan}}, \bibinfo {author} {\bibfnamefont {S.~M.}\ \bibnamefont {Lloyd}}, \
  and\ \bibinfo {author} {\bibfnamefont {M.}~\bibnamefont {Babiker}},\
  }\bibfield  {title} {\enquote {\bibinfo {title} {Chiral-specific
  electron-vortex-beam spectroscopy},}\ }\href {\doibase
  10.1103/PhysRevA.88.031801} {\bibfield  {journal} {\bibinfo  {journal} {Phys.
  Rev. A}\ }\textbf {\bibinfo {volume} {88}},\ \bibinfo {pages} {031801(R)}
  (\bibinfo {year} {2013})}\BibitemShut {NoStop}%
\bibitem [{\citenamefont {Pohl}\ \emph {et~al.}(2015)\citenamefont {Pohl},
  \citenamefont {Schneider}, \citenamefont {Rusz},\ and\ \citenamefont
  {Rellinghaus}}]{pohl2015measureEMCD}%
  \BibitemOpen
  \bibfield  {author} {\bibinfo {author} {\bibfnamefont {D.}~\bibnamefont
  {Pohl}}, \bibinfo {author} {\bibfnamefont {S.}~\bibnamefont {Schneider}},
  \bibinfo {author} {\bibfnamefont {J.}~\bibnamefont {Rusz}}, \ and\ \bibinfo
  {author} {\bibfnamefont {B.}~\bibnamefont {Rellinghaus}},\ }\bibfield
  {title} {\enquote {\bibinfo {title} {Electron vortex beams prepared by a
  spiral aperture with the goal to measure {EMCD} on ferromagnetic films via
  {STEM}},}\ }\href {\doibase 10.1016/j.ultramic.2014.11.025} {\bibfield
  {journal} {\bibinfo  {journal} {Ultramicroscopy}\ }\textbf {\bibinfo {volume}
  {150}},\ \bibinfo {pages} {16 -- 22} (\bibinfo {year} {2015})}\BibitemShut
  {NoStop}%
\bibitem [{\citenamefont {Schattschneider}\ \emph
  {et~al.}(2014{\natexlab{a}})\citenamefont {Schattschneider}, \citenamefont
  {L{\"o}ffler}, \citenamefont {St{\"o}ger-Pollach},\ and\ \citenamefont
  {Verbeeck}}]{schattschneider2014dichroismFeasible}%
  \BibitemOpen
  \bibfield  {author} {\bibinfo {author} {\bibfnamefont {P.}~\bibnamefont
  {Schattschneider}}, \bibinfo {author} {\bibfnamefont {S.}~\bibnamefont
  {L{\"o}ffler}}, \bibinfo {author} {\bibfnamefont {M.}~\bibnamefont
  {St{\"o}ger-Pollach}}, \ and\ \bibinfo {author} {\bibfnamefont
  {J.}~\bibnamefont {Verbeeck}},\ }\bibfield  {title} {\enquote {\bibinfo
  {title} {Is magnetic chiral dichroism feasible with electron vortices?}}\
  }\href {\doibase 10.1016/j.ultramic.2013.07.012} {\bibfield  {journal}
  {\bibinfo  {journal} {Ultramicroscopy}\ }\textbf {\bibinfo {volume} {136}},\
  \bibinfo {pages} {81 -- 85} (\bibinfo {year}
  {2014}{\natexlab{a}})}\BibitemShut {NoStop}%
\bibitem [{\citenamefont {Schachinger}\ \emph {et~al.}(2017)\citenamefont
  {Schachinger}, \citenamefont {L{\"o}ffler}, \citenamefont
  {Steiger-Thirsfeld}, \citenamefont {St{\"o}ger-Pollach}, \citenamefont
  {Schneider}, \citenamefont {Pohl}, \citenamefont {Rellinghaus},\ and\
  \citenamefont {Schattschneider}}]{schachinger2017EMCD}%
  \BibitemOpen
  \bibfield  {author} {\bibinfo {author} {\bibfnamefont {T.}~\bibnamefont
  {Schachinger}}, \bibinfo {author} {\bibfnamefont {S.}~\bibnamefont
  {L{\"o}ffler}}, \bibinfo {author} {\bibfnamefont {A.}~\bibnamefont
  {Steiger-Thirsfeld}}, \bibinfo {author} {\bibfnamefont {M.}~\bibnamefont
  {St{\"o}ger-Pollach}}, \bibinfo {author} {\bibfnamefont {S.}~\bibnamefont
  {Schneider}}, \bibinfo {author} {\bibfnamefont {D.}~\bibnamefont {Pohl}},
  \bibinfo {author} {\bibfnamefont {B.}~\bibnamefont {Rellinghaus}}, \ and\
  \bibinfo {author} {\bibfnamefont {P.}~\bibnamefont {Schattschneider}},\
  }\bibfield  {title} {\enquote {\bibinfo {title} {{EMCD} with an electron
  vortex filter: Limitations and possibilities},}\ }\href {\doibase
  10.1016/j.ultramic.2017.03.019} {\bibfield  {journal} {\bibinfo  {journal}
  {Ultramicroscopy}\ }\textbf {\bibinfo {volume} {179}},\ \bibinfo {pages} {15
  -- 23} (\bibinfo {year} {2017})}\BibitemShut {NoStop}%
\bibitem [{\citenamefont {Edstr\"om}\ \emph
  {et~al.}(2016{\natexlab{a}})\citenamefont {Edstr\"om}, \citenamefont {Lubk},\
  and\ \citenamefont {Rusz}}]{edstrom2016scatteringPRL}%
  \BibitemOpen
  \bibfield  {author} {\bibinfo {author} {\bibfnamefont {A.}~\bibnamefont
  {Edstr\"om}}, \bibinfo {author} {\bibfnamefont {A.}~\bibnamefont {Lubk}}, \
  and\ \bibinfo {author} {\bibfnamefont {J.}~\bibnamefont {Rusz}},\ }\bibfield
  {title} {\enquote {\bibinfo {title} {Elastic scattering of electron vortex
  beams in magnetic matter},}\ }\href {\doibase 10.1103/PhysRevLett.116.127203}
  {\bibfield  {journal} {\bibinfo  {journal} {Phys. Rev. Lett.}\ }\textbf
  {\bibinfo {volume} {116}},\ \bibinfo {pages} {127203} (\bibinfo {year}
  {2016}{\natexlab{a}})}\BibitemShut {NoStop}%
\bibitem [{\citenamefont {Edstr\"om}\ \emph
  {et~al.}(2016{\natexlab{b}})\citenamefont {Edstr\"om}, \citenamefont {Lubk},\
  and\ \citenamefont {Rusz}}]{edstrom2016scattering}%
  \BibitemOpen
  \bibfield  {author} {\bibinfo {author} {\bibfnamefont {A.}~\bibnamefont
  {Edstr\"om}}, \bibinfo {author} {\bibfnamefont {A.}~\bibnamefont {Lubk}}, \
  and\ \bibinfo {author} {\bibfnamefont {J.}~\bibnamefont {Rusz}},\ }\bibfield
  {title} {\enquote {\bibinfo {title} {Magnetic effects in the paraxial regime
  of elastic electron scattering},}\ }\href {\doibase
  10.1103/PhysRevB.94.174414} {\bibfield  {journal} {\bibinfo  {journal} {Phys.
  Rev. B}\ }\textbf {\bibinfo {volume} {94}},\ \bibinfo {pages} {174414}
  (\bibinfo {year} {2016}{\natexlab{b}})}\BibitemShut {NoStop}%
\bibitem [{\citenamefont {Gallatin}\ and\ \citenamefont
  {McMorran}(2012)}]{gallatin2012propagation}%
  \BibitemOpen
  \bibfield  {author} {\bibinfo {author} {\bibfnamefont {G.~M.}\ \bibnamefont
  {Gallatin}}\ and\ \bibinfo {author} {\bibfnamefont {B.}~\bibnamefont
  {McMorran}},\ }\bibfield  {title} {\enquote {\bibinfo {title} {Propagation of
  vortex electron wave functions in a magnetic field},}\ }\href {\doibase
  10.1103/PhysRevA.86.012701} {\bibfield  {journal} {\bibinfo  {journal} {Phys.
  Rev. A}\ }\textbf {\bibinfo {volume} {86}},\ \bibinfo {pages} {012701}
  (\bibinfo {year} {2012})}\BibitemShut {NoStop}%
\bibitem [{\citenamefont {Greenshields}\ \emph {et~al.}(2012)\citenamefont
  {Greenshields}, \citenamefont {Stamps},\ and\ \citenamefont
  {Franke-Arnold}}]{greenshields2012vacuum}%
  \BibitemOpen
  \bibfield  {author} {\bibinfo {author} {\bibfnamefont {C.~R.}\ \bibnamefont
  {Greenshields}}, \bibinfo {author} {\bibfnamefont {R.~L.}\ \bibnamefont
  {Stamps}}, \ and\ \bibinfo {author} {\bibfnamefont {S.}~\bibnamefont
  {Franke-Arnold}},\ }\bibfield  {title} {\enquote {\bibinfo {title} {Vacuum
  {F}araday effect for electrons},}\ }\href {\doibase
  10.1088/1367-2630/14/10/103040} {\bibfield  {journal} {\bibinfo  {journal}
  {New Journal of Physics}\ }\textbf {\bibinfo {volume} {14}},\ \bibinfo
  {pages} {103040} (\bibinfo {year} {2012})}\BibitemShut {NoStop}%
\bibitem [{\citenamefont {Karimi}\ \emph {et~al.}(2012)\citenamefont {Karimi},
  \citenamefont {Marrucci}, \citenamefont {Grillo},\ and\ \citenamefont
  {Santamato}}]{karimi2012conversion}%
  \BibitemOpen
  \bibfield  {author} {\bibinfo {author} {\bibfnamefont {E.}~\bibnamefont
  {Karimi}}, \bibinfo {author} {\bibfnamefont {L.}~\bibnamefont {Marrucci}},
  \bibinfo {author} {\bibfnamefont {V.}~\bibnamefont {Grillo}}, \ and\ \bibinfo
  {author} {\bibfnamefont {E.}~\bibnamefont {Santamato}},\ }\bibfield  {title}
  {\enquote {\bibinfo {title} {Spin-to-orbital angular momentum conversion and
  spin-polarization filtering in electron beams},}\ }\href {\doibase
  10.1103/PhysRevLett.108.044801} {\bibfield  {journal} {\bibinfo  {journal}
  {Phys. Rev. Lett.}\ }\textbf {\bibinfo {volume} {108}},\ \bibinfo {pages}
  {044801} (\bibinfo {year} {2012})}\BibitemShut {NoStop}%
\bibitem [{\citenamefont {Littlejohn}\ and\ \citenamefont
  {Weigert}(1993)}]{littlejohn1993neutral}%
  \BibitemOpen
  \bibfield  {author} {\bibinfo {author} {\bibfnamefont {R.~G.}\ \bibnamefont
  {Littlejohn}}\ and\ \bibinfo {author} {\bibfnamefont {S.}~\bibnamefont
  {Weigert}},\ }\bibfield  {title} {\enquote {\bibinfo {title} {Adiabatic
  motion of a neutral spinning particle in an inhomogeneous magnetic field},}\
  }\href {\doibase 10.1103/PhysRevA.48.924} {\bibfield  {journal} {\bibinfo
  {journal} {Phys. Rev. A}\ }\textbf {\bibinfo {volume} {48}},\ \bibinfo
  {pages} {924--940} (\bibinfo {year} {1993})}\BibitemShut {NoStop}%
\bibitem [{\citenamefont {Aharonov}\ and\ \citenamefont
  {Stern}(1992)}]{aharanov1992berry}%
  \BibitemOpen
  \bibfield  {author} {\bibinfo {author} {\bibfnamefont {Y.}~\bibnamefont
  {Aharonov}}\ and\ \bibinfo {author} {\bibfnamefont {A.}~\bibnamefont
  {Stern}},\ }\bibfield  {title} {\enquote {\bibinfo {title} {Origin of the
  geometric forces accompanying {B}erry's geometric potentials},}\ }\href
  {\doibase 10.1103/PhysRevLett.69.3593} {\bibfield  {journal} {\bibinfo
  {journal} {Phys. Rev. Lett.}\ }\textbf {\bibinfo {volume} {69}},\ \bibinfo
  {pages} {3593--3597} (\bibinfo {year} {1992})}\BibitemShut {NoStop}%
\bibitem [{\citenamefont {Guzzinati}\ \emph {et~al.}(2013)\citenamefont
  {Guzzinati}, \citenamefont {Schattschneider}, \citenamefont {Bliokh},
  \citenamefont {Nori},\ and\ \citenamefont
  {Verbeeck}}]{guzzinati2013observation}%
  \BibitemOpen
  \bibfield  {author} {\bibinfo {author} {\bibfnamefont {G.}~\bibnamefont
  {Guzzinati}}, \bibinfo {author} {\bibfnamefont {P.}~\bibnamefont
  {Schattschneider}}, \bibinfo {author} {\bibfnamefont {K.~Y.}\ \bibnamefont
  {Bliokh}}, \bibinfo {author} {\bibfnamefont {F.}~\bibnamefont {Nori}}, \ and\
  \bibinfo {author} {\bibfnamefont {J.}~\bibnamefont {Verbeeck}},\ }\bibfield
  {title} {\enquote {\bibinfo {title} {Observation of the {L}armor and {G}ouy
  rotations with electron vortex beams},}\ }\href {\doibase
  10.1103/PhysRevLett.110.093601} {\bibfield  {journal} {\bibinfo  {journal}
  {Phys. Rev. Lett.}\ }\textbf {\bibinfo {volume} {110}},\ \bibinfo {pages}
  {093601} (\bibinfo {year} {2013})}\BibitemShut {NoStop}%
\bibitem [{\citenamefont {Lubk}\ \emph {et~al.}(2013)\citenamefont {Lubk},
  \citenamefont {Guzzinati}, \citenamefont {B\"orrnert},\ and\ \citenamefont
  {Verbeeck}}]{lubk2013transportPhase}%
  \BibitemOpen
  \bibfield  {author} {\bibinfo {author} {\bibfnamefont {A.}~\bibnamefont
  {Lubk}}, \bibinfo {author} {\bibfnamefont {G.}~\bibnamefont {Guzzinati}},
  \bibinfo {author} {\bibfnamefont {F.}~\bibnamefont {B\"orrnert}}, \ and\
  \bibinfo {author} {\bibfnamefont {J.}~\bibnamefont {Verbeeck}},\ }\bibfield
  {title} {\enquote {\bibinfo {title} {Transport of intensity phase retrieval
  of arbitrary wave fields including vortices},}\ }\href {\doibase
  10.1103/PhysRevLett.111.173902} {\bibfield  {journal} {\bibinfo  {journal}
  {Phys. Rev. Lett.}\ }\textbf {\bibinfo {volume} {111}},\ \bibinfo {pages}
  {173902} (\bibinfo {year} {2013})}\BibitemShut {NoStop}%
\bibitem [{\citenamefont {Allen}\ \emph {et~al.}(2001)\citenamefont {Allen},
  \citenamefont {Faulkner}, \citenamefont {Oxley},\ and\ \citenamefont
  {Paganin}}]{allen2001phaseRetrieval}%
  \BibitemOpen
  \bibfield  {author} {\bibinfo {author} {\bibfnamefont {L.~J.}\ \bibnamefont
  {Allen}}, \bibinfo {author} {\bibfnamefont {H.~M.~L.}\ \bibnamefont
  {Faulkner}}, \bibinfo {author} {\bibfnamefont {M.~P.}\ \bibnamefont {Oxley}},
  \ and\ \bibinfo {author} {\bibfnamefont {D.}~\bibnamefont {Paganin}},\
  }\bibfield  {title} {\enquote {\bibinfo {title} {Phase retrieval and
  aberration correction in the presence of vortices in high-resolution
  transmission electron microscopy},}\ }\href {\doibase
  10.1016/S0304-3991(01)00072-9} {\bibfield  {journal} {\bibinfo  {journal}
  {Ultramicroscopy}\ }\textbf {\bibinfo {volume} {88}},\ \bibinfo {pages} {85
  -- 97} (\bibinfo {year} {2001})}\BibitemShut {NoStop}%
\bibitem [{\citenamefont {B{\'e}ch{\'e}}\ \emph {et~al.}(2014)\citenamefont
  {B{\'e}ch{\'e}}, \citenamefont {Van~Boxem}, \citenamefont {Van~Tendeloo},\
  and\ \citenamefont {Verbeeck}}]{beche2014magnetic}%
  \BibitemOpen
  \bibfield  {author} {\bibinfo {author} {\bibfnamefont {A.}~\bibnamefont
  {B{\'e}ch{\'e}}}, \bibinfo {author} {\bibfnamefont {R.}~\bibnamefont
  {Van~Boxem}}, \bibinfo {author} {\bibfnamefont {G.}~\bibnamefont
  {Van~Tendeloo}}, \ and\ \bibinfo {author} {\bibfnamefont {J.}~\bibnamefont
  {Verbeeck}},\ }\bibfield  {title} {\enquote {\bibinfo {title} {Magnetic
  monopole field exposed by electrons},}\ }\href {\doibase 10.1038/nphys2816}
  {\bibfield  {journal} {\bibinfo  {journal} {Nature Physics}\ }\textbf
  {\bibinfo {volume} {10}},\ \bibinfo {pages} {26--29} (\bibinfo {year}
  {2014})}\BibitemShut {NoStop}%
\bibitem [{\citenamefont {Ivanov}\ \emph {et~al.}(2016)\citenamefont {Ivanov},
  \citenamefont {Seipt}, \citenamefont {Surzhykov},\ and\ \citenamefont
  {Fritzsche}}]{ivanov2016double}%
  \BibitemOpen
  \bibfield  {author} {\bibinfo {author} {\bibfnamefont {I.~P.}\ \bibnamefont
  {Ivanov}}, \bibinfo {author} {\bibfnamefont {D.}~\bibnamefont {Seipt}},
  \bibinfo {author} {\bibfnamefont {A.}~\bibnamefont {Surzhykov}}, \ and\
  \bibinfo {author} {\bibfnamefont {S.}~\bibnamefont {Fritzsche}},\ }\bibfield
  {title} {\enquote {\bibinfo {title} {Double-slit experiment in momentum
  space},}\ }\href {\doibase 10.1209/0295-5075/115/41001} {\bibfield  {journal}
  {\bibinfo  {journal} {EPL (Europhysics Letters)}\ }\textbf {\bibinfo {volume}
  {115}},\ \bibinfo {pages} {41001} (\bibinfo {year} {2016})}\BibitemShut
  {NoStop}%
\bibitem [{\citenamefont {Schattschneider}\ \emph
  {et~al.}(2014{\natexlab{b}})\citenamefont {Schattschneider}, \citenamefont
  {Schachinger}, \citenamefont {St{\"o}ger-Pollach}, \citenamefont
  {L{\"o}ffler}, \citenamefont {Steiger-Thirsfeld}, \citenamefont {Bliokh},\
  and\ \citenamefont {Nori}}]{schattschneider2014imaging}%
  \BibitemOpen
  \bibfield  {author} {\bibinfo {author} {\bibfnamefont {P.}~\bibnamefont
  {Schattschneider}}, \bibinfo {author} {\bibfnamefont {Th.}\ \bibnamefont
  {Schachinger}}, \bibinfo {author} {\bibfnamefont {M.}~\bibnamefont
  {St{\"o}ger-Pollach}}, \bibinfo {author} {\bibfnamefont {S.}~\bibnamefont
  {L{\"o}ffler}}, \bibinfo {author} {\bibfnamefont {A.}~\bibnamefont
  {Steiger-Thirsfeld}}, \bibinfo {author} {\bibfnamefont {K.~Y.}\ \bibnamefont
  {Bliokh}}, \ and\ \bibinfo {author} {\bibfnamefont {F.}~\bibnamefont
  {Nori}},\ }\bibfield  {title} {\enquote {\bibinfo {title} {Imaging the
  dynamics of free-electron {L}andau states},}\ }\href {\doibase
  10.1038/ncomms5586} {\bibfield  {journal} {\bibinfo  {journal} {Nature
  communications}\ }\textbf {\bibinfo {volume} {5}},\ \bibinfo {pages} {4586}
  (\bibinfo {year} {2014}{\natexlab{b}})}\BibitemShut {NoStop}%
\bibitem [{\citenamefont {Batelaan}\ \emph {et~al.}(1997)\citenamefont
  {Batelaan}, \citenamefont {Gay},\ and\ \citenamefont
  {Schwendiman}}]{batelaan1997sternGerlach}%
  \BibitemOpen
  \bibfield  {author} {\bibinfo {author} {\bibfnamefont {H.}~\bibnamefont
  {Batelaan}}, \bibinfo {author} {\bibfnamefont {T.~J.}\ \bibnamefont {Gay}}, \
  and\ \bibinfo {author} {\bibfnamefont {J.~J.}\ \bibnamefont {Schwendiman}},\
  }\bibfield  {title} {\enquote {\bibinfo {title} {{S}tern-{G}erlach effect for
  electron beams},}\ }\href {\doibase 10.1103/PhysRevLett.79.4517} {\bibfield
  {journal} {\bibinfo  {journal} {Phys. Rev. Lett.}\ }\textbf {\bibinfo
  {volume} {79}},\ \bibinfo {pages} {4517--4521} (\bibinfo {year}
  {1997})}\BibitemShut {NoStop}%
\bibitem [{\citenamefont {Gallup}\ \emph {et~al.}(2001)\citenamefont {Gallup},
  \citenamefont {Batelaan},\ and\ \citenamefont
  {Gay}}]{gallup2001sternGerlach}%
  \BibitemOpen
  \bibfield  {author} {\bibinfo {author} {\bibfnamefont {G.~A.}\ \bibnamefont
  {Gallup}}, \bibinfo {author} {\bibfnamefont {H.}~\bibnamefont {Batelaan}}, \
  and\ \bibinfo {author} {\bibfnamefont {T.~J.}\ \bibnamefont {Gay}},\
  }\bibfield  {title} {\enquote {\bibinfo {title} {Quantum-mechanical analysis
  of a longitudinal {S}tern-{G}erlach effect},}\ }\href {\doibase
  10.1103/PhysRevLett.86.4508} {\bibfield  {journal} {\bibinfo  {journal}
  {Phys. Rev. Lett.}\ }\textbf {\bibinfo {volume} {86}},\ \bibinfo {pages}
  {4508--4511} (\bibinfo {year} {2001})}\BibitemShut {NoStop}%
\bibitem [{\citenamefont {Harvey}\ \emph {et~al.}(2017)\citenamefont {Harvey},
  \citenamefont {Grillo},\ and\ \citenamefont
  {McMorran}}]{harvey2017sternExperimental}%
  \BibitemOpen
  \bibfield  {author} {\bibinfo {author} {\bibfnamefont {T.~R.}\ \bibnamefont
  {Harvey}}, \bibinfo {author} {\bibfnamefont {V.}~\bibnamefont {Grillo}}, \
  and\ \bibinfo {author} {\bibfnamefont {B.~J.}\ \bibnamefont {McMorran}},\
  }\bibfield  {title} {\enquote {\bibinfo {title} {{S}tern-{G}erlach-like
  approach to electron orbital angular momentum measurement},}\ }\href
  {\doibase 10.1103/PhysRevA.95.021801} {\bibfield  {journal} {\bibinfo
  {journal} {Phys. Rev. A}\ }\textbf {\bibinfo {volume} {95}},\ \bibinfo
  {pages} {021801(R)} (\bibinfo {year} {2017})}\BibitemShut {NoStop}%
\bibitem [{\citenamefont {Schattschneider}\ \emph {et~al.}(2017)\citenamefont
  {Schattschneider}, \citenamefont {Grillo},\ and\ \citenamefont
  {Aubry}}]{schattschneider2017spinPolarization}%
  \BibitemOpen
  \bibfield  {author} {\bibinfo {author} {\bibfnamefont {P.}~\bibnamefont
  {Schattschneider}}, \bibinfo {author} {\bibfnamefont {V.}~\bibnamefont
  {Grillo}}, \ and\ \bibinfo {author} {\bibfnamefont {D.}~\bibnamefont
  {Aubry}},\ }\bibfield  {title} {\enquote {\bibinfo {title} {Spin polarisation
  with electron {B}essel beams},}\ }\href {\doibase
  10.1016/j.ultramic.2016.11.029} {\bibfield  {journal} {\bibinfo  {journal}
  {Ultramicroscopy}\ }\textbf {\bibinfo {volume} {176}},\ \bibinfo {pages} {188
  -- 193} (\bibinfo {year} {2017})}\BibitemShut {NoStop}%
\bibitem [{\citenamefont {Karimi}\ \emph {et~al.}(2014)\citenamefont {Karimi},
  \citenamefont {Grillo}, \citenamefont {Boyd},\ and\ \citenamefont
  {Santamato}}]{karimi2014spinPolarized}%
  \BibitemOpen
  \bibfield  {author} {\bibinfo {author} {\bibfnamefont {E.}~\bibnamefont
  {Karimi}}, \bibinfo {author} {\bibfnamefont {V.}~\bibnamefont {Grillo}},
  \bibinfo {author} {\bibfnamefont {R.~W.}\ \bibnamefont {Boyd}}, \ and\
  \bibinfo {author} {\bibfnamefont {E.}~\bibnamefont {Santamato}},\ }\bibfield
  {title} {\enquote {\bibinfo {title} {Generation of a spin-polarized electron
  beam by multipole magnetic fields},}\ }\href {\doibase
  10.1016/j.ultramic.2013.12.002} {\bibfield  {journal} {\bibinfo  {journal}
  {Ultramicroscopy}\ }\textbf {\bibinfo {volume} {138}},\ \bibinfo {pages} {22
  -- 27} (\bibinfo {year} {2014})}\BibitemShut {NoStop}%
\bibitem [{\citenamefont {Grillo}\ \emph {et~al.}(2013)\citenamefont {Grillo},
  \citenamefont {Marrucci}, \citenamefont {Karimi}, \citenamefont {Zanella},\
  and\ \citenamefont {Santamato}}]{grillo2013spinDevice}%
  \BibitemOpen
  \bibfield  {author} {\bibinfo {author} {\bibfnamefont {V.}~\bibnamefont
  {Grillo}}, \bibinfo {author} {\bibfnamefont {L.}~\bibnamefont {Marrucci}},
  \bibinfo {author} {\bibfnamefont {E.}~\bibnamefont {Karimi}}, \bibinfo
  {author} {\bibfnamefont {R.}~\bibnamefont {Zanella}}, \ and\ \bibinfo
  {author} {\bibfnamefont {E.}~\bibnamefont {Santamato}},\ }\bibfield  {title}
  {\enquote {\bibinfo {title} {Quantum simulation of a spin polarization device
  in an electron microscope},}\ }\href {\doibase 10.1088/1367-2630/15/9/093026}
  {\bibfield  {journal} {\bibinfo  {journal} {New Journal of Physics}\ }\textbf
  {\bibinfo {volume} {15}},\ \bibinfo {pages} {093026} (\bibinfo {year}
  {2013})}\BibitemShut {NoStop}%
\bibitem [{\citenamefont {Stoler}(1981)}]{stoler1981operator}%
  \BibitemOpen
  \bibfield  {author} {\bibinfo {author} {\bibfnamefont {D.}~\bibnamefont
  {Stoler}},\ }\bibfield  {title} {\enquote {\bibinfo {title} {Operator methods
  in physical optics},}\ }\href {\doibase 10.1364/JOSA.71.000334} {\bibfield
  {journal} {\bibinfo  {journal} {J. Opt. Soc. Am.}\ }\textbf {\bibinfo
  {volume} {71}},\ \bibinfo {pages} {334--341} (\bibinfo {year}
  {1981})}\BibitemShut {NoStop}%
\bibitem [{\citenamefont {van Enk}\ and\ \citenamefont
  {Nienhuis}(1992)}]{vanEnk1992eigenfunction}%
  \BibitemOpen
  \bibfield  {author} {\bibinfo {author} {\bibfnamefont {S.~J.}\ \bibnamefont
  {van Enk}}\ and\ \bibinfo {author} {\bibfnamefont {G.}~\bibnamefont
  {Nienhuis}},\ }\bibfield  {title} {\enquote {\bibinfo {title} {Eigenfunction
  description of laser beams and orbital angular momentum of light},}\ }\href
  {\doibase 10.1016/0030-4018(92)90424-P} {\bibfield  {journal} {\bibinfo
  {journal} {Optics Communications}\ }\textbf {\bibinfo {volume} {94}},\
  \bibinfo {pages} {147--158} (\bibinfo {year} {1992})}\BibitemShut {NoStop}%
\bibitem [{\citenamefont {Szil{\'a}gyi}(1998)}]{szilagyi1998glaser}%
  \BibitemOpen
  \bibfield  {author} {\bibinfo {author} {\bibfnamefont {M.}~\bibnamefont
  {Szil{\'a}gyi}},\ }\href {\doibase 10.1007/978-1-4613-0923-9} {\emph
  {\bibinfo {title} {{E}lectron and {I}on {O}ptics}}}\ (\bibinfo  {publisher}
  {Springer US},\ \bibinfo {year} {1998})\BibitemShut {NoStop}%
\bibitem [{Note1()}]{Note1}%
  \BibitemOpen
  \bibinfo {note} {The extra radial term in the magnetic field is a consequence
  of the fact that $\protect \boldsymbol \nabla . \protect \boldsymbol B$ needs
  to be zero.}\BibitemShut {Stop}%
\bibitem [{\citenamefont {Kitadono}\ \emph {et~al.}(2020)\citenamefont
  {Kitadono}, \citenamefont {Wakamatsu}, \citenamefont {Zou},\ and\
  \citenamefont {Zhang}}]{kitadono2020guidingCenter}%
  \BibitemOpen
  \bibfield  {author} {\bibinfo {author} {\bibfnamefont {Y.}~\bibnamefont
  {Kitadono}}, \bibinfo {author} {\bibfnamefont {M.}~\bibnamefont {Wakamatsu}},
  \bibinfo {author} {\bibfnamefont {L.}~\bibnamefont {Zou}}, \ and\ \bibinfo
  {author} {\bibfnamefont {P.}~\bibnamefont {Zhang}},\ }\bibfield  {title}
  {\enquote {\bibinfo {title} {Role of guiding center in {L}andau level system
  and mechanical and pseudo orbital angular momenta},}\ }\href {\doibase
  10.1142/S0217751X20500967} {\bibfield  {journal} {\bibinfo  {journal}
  {International Journal of Modern Physics A}\ }\textbf {\bibinfo {volume}
  {35}},\ \bibinfo {pages} {2050096} (\bibinfo {year} {2020})}\BibitemShut
  {NoStop}%
\bibitem [{\citenamefont {van Enk}(2020)}]{vanEnk2020angular}%
  \BibitemOpen
  \bibfield  {author} {\bibinfo {author} {\bibfnamefont {S.~J.}\ \bibnamefont
  {van Enk}},\ }\bibfield  {title} {\enquote {\bibinfo {title} {Angular
  momentum in the fractional quantum hall effect},}\ }\href {\doibase
  10.1119/10.0000831} {\bibfield  {journal} {\bibinfo  {journal} {American
  Journal of Physics}\ }\textbf {\bibinfo {volume} {88}},\ \bibinfo {pages}
  {286--291} (\bibinfo {year} {2020})}\BibitemShut {NoStop}%
\bibitem [{\citenamefont {Wakamatsu}\ \emph {et~al.}(2020)\citenamefont
  {Wakamatsu}, \citenamefont {Kitadono}, \citenamefont {Zou},\ and\
  \citenamefont {Zhang}}]{wakamatsu2020gauge}%
  \BibitemOpen
  \bibfield  {author} {\bibinfo {author} {\bibfnamefont {M.}~\bibnamefont
  {Wakamatsu}}, \bibinfo {author} {\bibfnamefont {Y.}~\bibnamefont {Kitadono}},
  \bibinfo {author} {\bibfnamefont {L.}~\bibnamefont {Zou}}, \ and\ \bibinfo
  {author} {\bibfnamefont {P.}~\bibnamefont {Zhang}},\ }\bibfield  {title}
  {\enquote {\bibinfo {title} {The physics of helical electron beam in a
  uniform magnetic field as a testing ground of gauge principle},}\ }\href
  {\doibase 10.1016/j.physleta.2020.126415} {\bibfield  {journal} {\bibinfo
  {journal} {Physics Letters A}\ }\textbf {\bibinfo {volume} {384}},\ \bibinfo
  {pages} {126415} (\bibinfo {year} {2020})}\BibitemShut {NoStop}%
\bibitem [{\citenamefont {Cohen-Tannoudji}\ \emph {et~al.}(1992)\citenamefont
  {Cohen-Tannoudji}, \citenamefont {Dupont-Roc}, \citenamefont {Grynberg},\
  and\ \citenamefont {Scully}}]{cohen1997electrodynamics}%
  \BibitemOpen
  \bibfield  {author} {\bibinfo {author} {\bibfnamefont {C.}~\bibnamefont
  {Cohen-Tannoudji}}, \bibinfo {author} {\bibfnamefont {J.}~\bibnamefont
  {Dupont-Roc}}, \bibinfo {author} {\bibfnamefont {G.}~\bibnamefont
  {Grynberg}}, \ and\ \bibinfo {author} {\bibfnamefont {M.~O.}\ \bibnamefont
  {Scully}},\ }\href {\doibase 10.1002/9783527618422} {\emph {\bibinfo {title}
  {Photons \& {A}toms - {I}ntroduction to {Q}uantum {E}lectrodynamics}}}\
  (\bibinfo  {publisher} {Wiley-VCH},\ \bibinfo {year} {1992})\BibitemShut
  {NoStop}%
\bibitem [{\citenamefont {Greenshields}\ \emph {et~al.}(2014)\citenamefont
  {Greenshields}, \citenamefont {Stamps}, \citenamefont {Franke-Arnold},\ and\
  \citenamefont {Barnett}}]{greenshields2014conserved}%
  \BibitemOpen
  \bibfield  {author} {\bibinfo {author} {\bibfnamefont {C.~R.}\ \bibnamefont
  {Greenshields}}, \bibinfo {author} {\bibfnamefont {R.~L.}\ \bibnamefont
  {Stamps}}, \bibinfo {author} {\bibfnamefont {S.}~\bibnamefont
  {Franke-Arnold}}, \ and\ \bibinfo {author} {\bibfnamefont {S.~M.}\
  \bibnamefont {Barnett}},\ }\bibfield  {title} {\enquote {\bibinfo {title} {Is
  the angular momentum of an electron conserved in a uniform magnetic field?}}\
  }\href {\doibase 10.1103/PhysRevLett.113.240404} {\bibfield  {journal}
  {\bibinfo  {journal} {Phys. Rev. Lett.}\ }\textbf {\bibinfo {volume} {113}},\
  \bibinfo {pages} {240404} (\bibinfo {year} {2014})}\BibitemShut {NoStop}%
\bibitem [{\citenamefont {Lewis}\ and\ \citenamefont
  {Riesenfeld}(1969)}]{lewis1969conserved}%
  \BibitemOpen
  \bibfield  {author} {\bibinfo {author} {\bibfnamefont {H.~R.}\ \bibnamefont
  {Lewis}}\ and\ \bibinfo {author} {\bibfnamefont {W.~B.}\ \bibnamefont
  {Riesenfeld}},\ }\bibfield  {title} {\enquote {\bibinfo {title} {An exact
  quantum theory of the time‐dependent harmonic oscillator and of a charged
  particle in a time‐dependent electromagnetic field},}\ }\href {\doibase
  10.1063/1.1664991} {\bibfield  {journal} {\bibinfo  {journal} {Journal of
  Mathematical Physics}\ }\textbf {\bibinfo {volume} {10}},\ \bibinfo {pages}
  {1458--1473} (\bibinfo {year} {1969})}\BibitemShut {NoStop}%
\bibitem [{\citenamefont {Leach}\ and\ \citenamefont
  {Andriopoulos}(2008)}]{leach2008ermakov}%
  \BibitemOpen
  \bibfield  {author} {\bibinfo {author} {\bibfnamefont {P.~G.~L.}\
  \bibnamefont {Leach}}\ and\ \bibinfo {author} {\bibfnamefont
  {K.}~\bibnamefont {Andriopoulos}},\ }\bibfield  {title} {\enquote {\bibinfo
  {title} {The {E}rmakov equation: a commentary},}\ }\href {\doibase
  10.2298/AADM0802146L} {\bibfield  {journal} {\bibinfo  {journal} {Applicable
  Analysis and Discrete Mathematics}\ ,\ \bibinfo {pages} {146--157}} (\bibinfo
  {year} {2008})}\BibitemShut {NoStop}%
\bibitem [{\citenamefont {Pinney}(1950)}]{pinney1950nonlinear}%
  \BibitemOpen
  \bibfield  {author} {\bibinfo {author} {\bibfnamefont {E.}~\bibnamefont
  {Pinney}},\ }\bibfield  {title} {\enquote {\bibinfo {title} {The nonlinear
  differential equation $y'' + p(x)y + cy^{-3} = 0$},}\ }\href {\doibase
  10.1090/S0002-9939-1950-0037979-4} {\bibfield  {journal} {\bibinfo  {journal}
  {Proceedings of the American Mathematical Society}\ }\textbf {\bibinfo
  {volume} {1}},\ \bibinfo {pages} {681} (\bibinfo {year} {1950})}\BibitemShut
  {NoStop}%
\bibitem [{\citenamefont {Allen}\ \emph {et~al.}(1999)\citenamefont {Allen},
  \citenamefont {Padgett},\ and\ \citenamefont
  {Babiker}}]{allen1999freeSpaceSolution}%
  \BibitemOpen
  \bibfield  {author} {\bibinfo {author} {\bibfnamefont {L.}~\bibnamefont
  {Allen}}, \bibinfo {author} {\bibfnamefont {M.~J.}\ \bibnamefont {Padgett}},
  \ and\ \bibinfo {author} {\bibfnamefont {M.}~\bibnamefont {Babiker}},\ }\href
  {\doibase 10.1016/S0079-6638(08)70391-3} {\emph {\bibinfo {title} {{IV} {T}he
  {O}rbital {A}ngular {M}omentum of {L}ight}}},\ edited by\ \bibinfo {editor}
  {\bibfnamefont {E.}~\bibnamefont {Wolf}},\ \bibinfo {series} {Progress in
  Optics}, Vol.~\bibinfo {volume} {39}\ (\bibinfo  {publisher} {Elsevier},\
  \bibinfo {year} {1999})\ pp.\ \bibinfo {pages} {291 -- 372}\BibitemShut
  {NoStop}%
\bibitem [{\citenamefont {Menouar}\ \emph {et~al.}(2010)\citenamefont
  {Menouar}, \citenamefont {Maamache},\ and\ \citenamefont
  {Choi}}]{menouar2010wavefunction}%
  \BibitemOpen
  \bibfield  {author} {\bibinfo {author} {\bibfnamefont {S.}~\bibnamefont
  {Menouar}}, \bibinfo {author} {\bibfnamefont {M.}~\bibnamefont {Maamache}}, \
  and\ \bibinfo {author} {\bibfnamefont {J.~R.}\ \bibnamefont {Choi}},\
  }\bibfield  {title} {\enquote {\bibinfo {title} {The time-dependent coupled
  oscillator model for the motion of a charged particle in the presence of a
  time-varying magnetic field},}\ }\href {\doibase
  10.1088/0031-8949/82/06/065004} {\bibfield  {journal} {\bibinfo  {journal}
  {Physica Scripta}\ }\textbf {\bibinfo {volume} {82}},\ \bibinfo {pages}
  {065004} (\bibinfo {year} {2010})}\BibitemShut {NoStop}%
\bibitem [{\citenamefont {Bliokh}\ \emph {et~al.}(2012)\citenamefont {Bliokh},
  \citenamefont {Schattschneider}, \citenamefont {Verbeeck},\ and\
  \citenamefont {Nori}}]{bliokh2012twist}%
  \BibitemOpen
  \bibfield  {author} {\bibinfo {author} {\bibfnamefont {K.~Y.}\ \bibnamefont
  {Bliokh}}, \bibinfo {author} {\bibfnamefont {P.}~\bibnamefont
  {Schattschneider}}, \bibinfo {author} {\bibfnamefont {J.}~\bibnamefont
  {Verbeeck}}, \ and\ \bibinfo {author} {\bibfnamefont {F.}~\bibnamefont
  {Nori}},\ }\bibfield  {title} {\enquote {\bibinfo {title} {Electron vortex
  beams in a magnetic field: A new twist on {L}andau levels and
  {A}haronov-{B}ohm states},}\ }\href {\doibase 10.1103/PhysRevX.2.041011}
  {\bibfield  {journal} {\bibinfo  {journal} {Phys. Rev. X}\ }\textbf {\bibinfo
  {volume} {2}},\ \bibinfo {pages} {041011} (\bibinfo {year}
  {2012})}\BibitemShut {NoStop}%
\bibitem [{\citenamefont {Lewis}(1968)}]{lewis1968invariant}%
  \BibitemOpen
  \bibfield  {author} {\bibinfo {author} {\bibfnamefont {H.~R.}\ \bibnamefont
  {Lewis}},\ }\bibfield  {title} {\enquote {\bibinfo {title} {Class of exact
  invariants for classical and quantum time‐dependent harmonic
  oscillators},}\ }\href {\doibase 10.1063/1.1664532} {\bibfield  {journal}
  {\bibinfo  {journal} {Journal of Mathematical Physics}\ }\textbf {\bibinfo
  {volume} {9}},\ \bibinfo {pages} {1976--1986} (\bibinfo {year}
  {1968})}\BibitemShut {NoStop}%
\bibitem [{\citenamefont {Eliezer}\ and\ \citenamefont
  {Gray}(1976)}]{eliezer1976note}%
  \BibitemOpen
  \bibfield  {author} {\bibinfo {author} {\bibfnamefont {C.~J.}\ \bibnamefont
  {Eliezer}}\ and\ \bibinfo {author} {\bibfnamefont {A.}~\bibnamefont {Gray}},\
  }\bibfield  {title} {\enquote {\bibinfo {title} {A note on the time-dependent
  harmonic oscillator},}\ }\href {\doibase doi.org/10.1137/0130043} {\bibfield
  {journal} {\bibinfo  {journal} {SIAM Journal on Applied Mathematics}\
  }\textbf {\bibinfo {volume} {30}},\ \bibinfo {pages} {463--468} (\bibinfo
  {year} {1976})}\BibitemShut {NoStop}%
\bibitem [{\citenamefont {{Landau}}(1930)}]{landau1930original}%
  \BibitemOpen
  \bibfield  {author} {\bibinfo {author} {\bibfnamefont {L.}~\bibnamefont
  {{Landau}}},\ }\bibfield  {title} {\enquote {\bibinfo {title}
  {{{D}iamagnetismus der {M}etalle}},}\ }\href {\doibase 10.1007/BF01397213}
  {\bibfield  {journal} {\bibinfo  {journal} {{Z}eitschrift f\"ur {P}hysik}\
  }\textbf {\bibinfo {volume} {64}},\ \bibinfo {pages} {629--637} (\bibinfo
  {year} {1930})}\BibitemShut {NoStop}%
\bibitem [{\citenamefont {Bliokh}\ \emph {et~al.}(2017)\citenamefont {Bliokh},
  \citenamefont {Ivanov}, \citenamefont {Guzzinati}, \citenamefont {Clark},
  \citenamefont {{Van Boxem}}, \citenamefont {B{\'e}ch{\'e}}, \citenamefont
  {Juchtmans}, \citenamefont {Alonso}, \citenamefont {Schattschneider},
  \citenamefont {Nori},\ and\ \citenamefont {Verbeeck}}]{bliokh2017theory}%
  \BibitemOpen
  \bibfield  {author} {\bibinfo {author} {\bibfnamefont {K.~Y.}\ \bibnamefont
  {Bliokh}}, \bibinfo {author} {\bibfnamefont {I.~P.}\ \bibnamefont {Ivanov}},
  \bibinfo {author} {\bibfnamefont {G.}~\bibnamefont {Guzzinati}}, \bibinfo
  {author} {\bibfnamefont {L.}~\bibnamefont {Clark}}, \bibinfo {author}
  {\bibfnamefont {R.}~\bibnamefont {{Van Boxem}}}, \bibinfo {author}
  {\bibfnamefont {A.}~\bibnamefont {B{\'e}ch{\'e}}}, \bibinfo {author}
  {\bibfnamefont {R.}~\bibnamefont {Juchtmans}}, \bibinfo {author}
  {\bibfnamefont {M.~A.}\ \bibnamefont {Alonso}}, \bibinfo {author}
  {\bibfnamefont {P.}~\bibnamefont {Schattschneider}}, \bibinfo {author}
  {\bibfnamefont {F.}~\bibnamefont {Nori}}, \ and\ \bibinfo {author}
  {\bibfnamefont {J.}~\bibnamefont {Verbeeck}},\ }\bibfield  {title} {\enquote
  {\bibinfo {title} {Theory and applications of free-electron vortex states},}\
  }\href {\doibase 10.1016/j.physrep.2017.05.006} {\bibfield  {journal}
  {\bibinfo  {journal} {Physics Reports}\ }\textbf {\bibinfo {volume} {690}},\
  \bibinfo {pages} {1 -- 70} (\bibinfo {year} {2017})}\BibitemShut {NoStop}%
\bibitem [{\citenamefont {Rechtsman}\ \emph {et~al.}(2013)\citenamefont
  {Rechtsman}, \citenamefont {Zeuner}, \citenamefont {Plotnik}, \citenamefont
  {Lumer}, \citenamefont {Podolsky}, \citenamefont {Dreisow}, \citenamefont
  {Nolte}, \citenamefont {Segev},\ and\ \citenamefont
  {Szameit}}]{rechtsman2013photonic}%
  \BibitemOpen
  \bibfield  {author} {\bibinfo {author} {\bibfnamefont {M.~C.}\ \bibnamefont
  {Rechtsman}}, \bibinfo {author} {\bibfnamefont {J.~M.}\ \bibnamefont
  {Zeuner}}, \bibinfo {author} {\bibfnamefont {Y.}~\bibnamefont {Plotnik}},
  \bibinfo {author} {\bibfnamefont {Y.}~\bibnamefont {Lumer}}, \bibinfo
  {author} {\bibfnamefont {D.}~\bibnamefont {Podolsky}}, \bibinfo {author}
  {\bibfnamefont {F.}~\bibnamefont {Dreisow}}, \bibinfo {author} {\bibfnamefont
  {S.}~\bibnamefont {Nolte}}, \bibinfo {author} {\bibfnamefont
  {M.}~\bibnamefont {Segev}}, \ and\ \bibinfo {author} {\bibfnamefont
  {A.}~\bibnamefont {Szameit}},\ }\bibfield  {title} {\enquote {\bibinfo
  {title} {Photonic {F}loquet topological insulators},}\ }\href {\doibase
  10.1038/nature12066} {\bibfield  {journal} {\bibinfo  {journal} {Nature}\
  }\textbf {\bibinfo {volume} {496}},\ \bibinfo {pages} {196--200} (\bibinfo
  {year} {2013})}\BibitemShut {NoStop}%
\bibitem [{\citenamefont {Lumer}\ \emph {et~al.}(2019)\citenamefont {Lumer},
  \citenamefont {Bandres}, \citenamefont {Heinrich}, \citenamefont {Maczewsky},
  \citenamefont {Herzig-Sheinfux}, \citenamefont {Szameit},\ and\ \citenamefont
  {Segev}}]{lumer2019lightGuiding}%
  \BibitemOpen
  \bibfield  {author} {\bibinfo {author} {\bibfnamefont {Y.}~\bibnamefont
  {Lumer}}, \bibinfo {author} {\bibfnamefont {M.~A.}\ \bibnamefont {Bandres}},
  \bibinfo {author} {\bibfnamefont {M.}~\bibnamefont {Heinrich}}, \bibinfo
  {author} {\bibfnamefont {L.~J.}\ \bibnamefont {Maczewsky}}, \bibinfo {author}
  {\bibfnamefont {H.}~\bibnamefont {Herzig-Sheinfux}}, \bibinfo {author}
  {\bibfnamefont {A.}~\bibnamefont {Szameit}}, \ and\ \bibinfo {author}
  {\bibfnamefont {M.}~\bibnamefont {Segev}},\ }\bibfield  {title} {\enquote
  {\bibinfo {title} {Light guiding by artificial gauge fields},}\ }\href
  {\doibase 10.1038/s41566-019-0370-1} {\bibfield  {journal} {\bibinfo
  {journal} {Nature Photonics}\ }\textbf {\bibinfo {volume} {13}},\ \bibinfo
  {pages} {339--345} (\bibinfo {year} {2019})}\BibitemShut {NoStop}%
\bibitem [{\citenamefont {Abbaszadeh}\ \emph {et~al.}(2017)\citenamefont
  {Abbaszadeh}, \citenamefont {Souslov}, \citenamefont {Paulose}, \citenamefont
  {Schomerus},\ and\ \citenamefont {Vitelli}}]{abbaszadeh2017sonicLandau}%
  \BibitemOpen
  \bibfield  {author} {\bibinfo {author} {\bibfnamefont {H.}~\bibnamefont
  {Abbaszadeh}}, \bibinfo {author} {\bibfnamefont {A.}~\bibnamefont {Souslov}},
  \bibinfo {author} {\bibfnamefont {J.}~\bibnamefont {Paulose}}, \bibinfo
  {author} {\bibfnamefont {H.}~\bibnamefont {Schomerus}}, \ and\ \bibinfo
  {author} {\bibfnamefont {V.}~\bibnamefont {Vitelli}},\ }\bibfield  {title}
  {\enquote {\bibinfo {title} {Sonic {L}andau levels and synthetic gauge fields
  in mechanical metamaterials},}\ }\href {\doibase
  10.1103/PhysRevLett.119.195502} {\bibfield  {journal} {\bibinfo  {journal}
  {Phys. Rev. Lett.}\ }\textbf {\bibinfo {volume} {119}},\ \bibinfo {pages}
  {195502} (\bibinfo {year} {2017})}\BibitemShut {NoStop}%
\bibitem [{\citenamefont {Dalibard}\ \emph {et~al.}(2011)\citenamefont
  {Dalibard}, \citenamefont {Gerbier}, \citenamefont {Juzeli\=unas},\ and\
  \citenamefont {\"Ohberg}}]{dalibard2011artificialGauge}%
  \BibitemOpen
  \bibfield  {author} {\bibinfo {author} {\bibfnamefont {J.}~\bibnamefont
  {Dalibard}}, \bibinfo {author} {\bibfnamefont {F.}~\bibnamefont {Gerbier}},
  \bibinfo {author} {\bibfnamefont {G.}~\bibnamefont {Juzeli\=unas}}, \ and\
  \bibinfo {author} {\bibfnamefont {P.}~\bibnamefont {\"Ohberg}},\ }\bibfield
  {title} {\enquote {\bibinfo {title} {Colloquium: Artificial gauge potentials
  for neutral atoms},}\ }\href {\doibase 10.1103/RevModPhys.83.1523} {\bibfield
   {journal} {\bibinfo  {journal} {Rev. Mod. Phys.}\ }\textbf {\bibinfo
  {volume} {83}},\ \bibinfo {pages} {1523--1543} (\bibinfo {year}
  {2011})}\BibitemShut {NoStop}%
\bibitem [{\citenamefont {Jagannathan}(1990)}]{jagannathan1990dirac}%
  \BibitemOpen
  \bibfield  {author} {\bibinfo {author} {\bibfnamefont {R.}~\bibnamefont
  {Jagannathan}},\ }\bibfield  {title} {\enquote {\bibinfo {title} {Quantum
  theory of electron lenses based on the {D}irac equation},}\ }\href {\doibase
  10.1103/PhysRevA.42.6674} {\bibfield  {journal} {\bibinfo  {journal} {Phys.
  Rev. A}\ }\textbf {\bibinfo {volume} {42}},\ \bibinfo {pages} {6674--6689}
  (\bibinfo {year} {1990})}\BibitemShut {NoStop}%
\end{thebibliography}%

\appendix
\section{Spin\label{Sec:Spin}}

An actual physical electron also has spin which interacts with the magnetic field via $H_B = - \bb \mu . \bb B \approx \frac{e}{m_e} \bb S . \bb B$ (using $g\approx 2$). In our case we have
\begin{equation}
	 H_B  \approx 2 S_z \Omega(z) - \rho \dot \Omega(z) S_\rho.
\end{equation}

While the first term is comparable in magnitude to the spinless Hamiltonian, the second term is small since by the assumption of paraxial optics, the beam is very close to the $z$ axis~\cite{jagannathan1990dirac}.

To incorporate spin one can construct two-component vector solutions from the scalar solutions we have provided~\cite{jagannathan1990dirac}. The first term can be dealt with exactly and will appear as an independent spectator degree of freedom without any effect on the dynamics. The second term, which is small and couples spin to spatial degrees of freedom, can be dealt with perturbatively.

\section{Solution of the lensing equation for the Glaser field}\label{Sec:Derivation}

To solve Eq. \eqref{lensing} for $\Omega(z) = \frac{\Omega_0}{1 + (z- c)^2/a^2}$ we first perform the transformation $x = z - c$ and write the equation as

\begin{equation}\label{Glasertosolve}
\frac{d^2w(x)}{dx^2} +\frac{\Omega_0^2a^4}{v^2} \frac{w(x)}{(a^2 + x^2)^2} - 
\frac{4}{k^2}\frac{1}{w^3(x)} = 0
\end{equation}

As shown in Ref.~\cite{pinney1950nonlinear}, we can first solve the linear differential equation

\begin{equation}
\frac{d^2w(x)}{dx^2} + \frac{\Omega_0^2a^4}{v^2} \frac{w(x)}{(a^2 + x^2)^2} = 0.
\end{equation}

to get two independent solutions

\begin{equation}
    y_1(x) = \sqrt{a^2 + x^2}\sin\bigg(\beta \arctan(x/a) \bigg)
\end{equation}

and

\begin{equation}
    y_2(x) = \sqrt{a^2 + x^2}\cos\bigg(\beta \arctan(x/a) \bigg)
\end{equation}

where $\beta = \sqrt{1 + a^2\Omega_0^2/v^2}$. Following the notation of Ref.~\cite{pinney1950nonlinear} we define $u(x) = \frac{w_c}{a}y_1(x) + \frac{w_c}{\beta R_c}y_2(x)$,  such that $u(0) = w_c$ and $\dot u(0) = \frac{w_c}{R_c}$, and $v(x) = y_2(x)$, so that $v(0) = 0$ and $\dot v(0) = \beta$. The Wronskian, $W$, of $u(x)$ and $v(x)$ is $u(x)\dot v(x) - \dot u(x) v(x) = w_c \beta$. Thus using Eq. 2 of Ref.~\cite{pinney1950nonlinear} the solution of the lensing equation for the Glaser field, Eq.\eqref{Glasertosolve}, is

\begin{align}
    w^2(x) = u^2(x) + \frac{4}{k^2W^2}v^2(x)
\end{align}

with $w(x = 0) = w_c$ and $\dot w(x = 0) = \frac{w_c}{R_c}$. The above solution can be simplified further to finally get Eq. \eqref{Glaserlensing}.

\end{document}